\documentclass[bibnotes,prl,twocolumn,superscriptaddress]{revtex4}%
\usepackage{epsfig,dsfont,amssymb,amsmath,amsthm,amsfonts,amsbsy,mathrsfs}
\usepackage{graphicx}
\usepackage{bm}%bold math
\begin{document}
\title{Radiation Pressure Cooling as a Quantum Dynamical Process}
\author{Bing He}
\email{binghe@uark.edu}
%\altaffiliation{Contributed equally to this work}
\affiliation{Department of Physics, University of Arkansas, Fayetteville, AR 72701, USA}
\author{Liu Yang}
%\altaffiliation{Contributed equally to this work}
\affiliation{College of Automation, Harbin Engineering University, Heilongjiang 150001, China}
\author{Qing Lin}
\affiliation{Department of Physics, University of Arkansas, Fayetteville, AR 72701, USA}
\affiliation{Fujian Provincial Key Laboratory of Light Propagation and Transformation, College of Information Science and Engineering, Huaqiao University, Xiamen 361021, China}
\author{Min Xiao}
\email{mxiao@uark.edu}
\affiliation{Department of Physics, University of Arkansas, Fayetteville, AR 72701, USA}
\affiliation{National Laboratory of Solid State Microstructures and School of Physics, Nanjing University, Nanjing 210093, China}

\begin{abstract}
One of the most fundamental problems in optomechanical cooling is how small the thermal phonon number of a mechanical oscillator can be achieved under the radiation pressure of a proper cavity field. Different from previous theoretical predictions, which were based on an optomechanical system's time-independent steady states, we treat such cooling as a dynamical process of driving the mechanical oscillator from its initial thermal state, due to its thermal equilibrium with the environment, to a stabilized quantum state of higher purity. We find that the stabilized thermal phonon number left in the end actually depends on how fast the cooling process could be. The cooling speed is decided by an effective optomechanical coupling intensity, which constitutes an essential parameter for cooling, in addition to the sideband resolution parameter that has been considered in other theoretical studies. The limiting thermal phonon number that any cooling process cannot surpass exhibits a discontinuous jump across a certain value of the parameter. 
\end{abstract}
\maketitle

Preparing the approximate pure quantum states of a sizable mechanical oscillator is a feasible way toward macroscopic quantumness. Practically starting from its thermal equilibrium with the environment, such process is implemented by coupling the oscillator to a cavity field generated by 
a red-detuned external drive, to reduce the associated thermal phonon number to a low level, similar to cooling the oscillator to a lower temperature. An important feature we will illustrate is that the cooling result depends on how fast the optomechanical system (OMS) evolves to the finally stable quantum state.

\begin{figure}[b!]
\vspace{0cm}
\centering
\epsfig{file=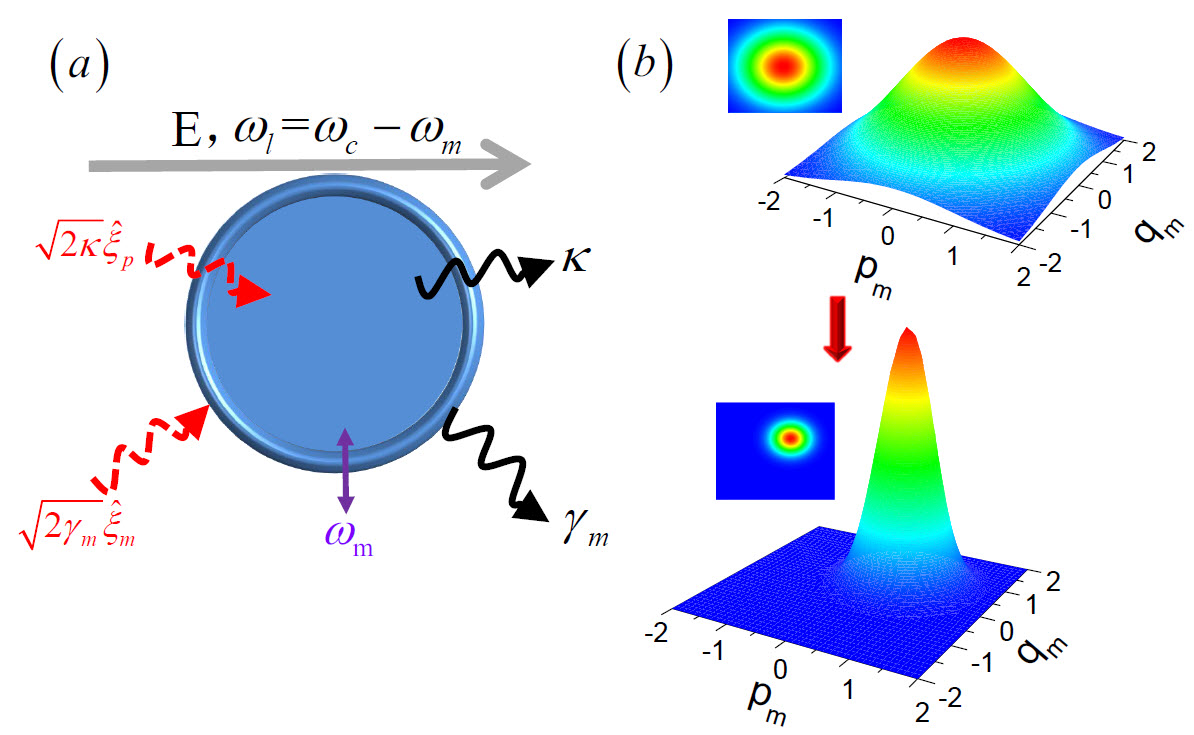,width=0.8\linewidth,clip=} 
{\vspace{-0.2cm}\caption{(a) A setup to perform optomechanical cooling. The quantum noises from the environment accompany 
the cavity and mechanical dissipations. (b) An ideal cooling process that transforms a thermal state (mixed) of the mechanical resonator to a coherent state (pure). The three-dimensional plots represent the corresponding Wigner functions. }}
\vspace{0cm}
\end{figure}

So far numerous experiments have realized the cooling to a few and even less than one mechanical quanta 
\cite{ex1,ex2,ex3,ex4,ex5,ex6,ex7,ex8,ex9, ex10, ex11, ex12, ex13}. Following the earlier study of quantum fluctuations under radiation pressure \cite{00, 01}, the theoretical description of such optomechanical cooling (see, e.g. \cite{c1,c2, c0,ck, c3,c4,c5,c6}) was based on a linearization procedure as that described in \cite{rev}; that is to decompose the cavity field mode $\hat{a}$ into the sum of the classical mean value $\alpha$ and its quantum fluctuation $\delta \hat{a}$. The linearized Hamiltonian gives the cooling action as a beamsplitter (BS) type coupling between the mechanical mode $\hat{b}$ and the fluctuation $\delta \hat{a}$ with their coupling intensity $g$ magnified by $\alpha$, which was generally treated as a constant of steady-state value. In an actual cooling process, however, the cavity mean field $\langle \hat{a}(t)\rangle=\alpha(t)$ is built up from zero (when the mechanical oscillator is in thermal equilibrium with its environment) and takes time to evolve to stable value. Then the effective coupling strength $g|\alpha|$ used in the previous studies should be more appropriately taken as a variable, since $\alpha(t)$ keeps changing during a cooling process. Due to the impossibility of finding the time-dependent $\alpha(t)$ analytically, it is difficult to study the cooling as a dynamical process if adopting the above-mentioned linearization. 

In the present work we put forward a quantum dynamical theory for optomechanical cooling. Using this completely quantum approach that linearizes a weakly coupled OMS's dynamics without resorting to its classical mean values such as $\alpha(t)$, one can numerically predict the involved physical quantities that evolve with time. The residual occupation of the thermal excitation of mechanical oscillator, as found in our approach, is connected with the whole cooling process. Compared with the previous theoretical predictions, this dynamical approach provides richer information about how good a quantum OMS can be fully cooled down to. 

Fig. 1(a) illustrates an exemplary optomechanical cooling setup. In a frame with the system modes rotating at the cavity frequency $\omega_c$ and the mechanical frequency $\omega_m$, respectively, the process is governed by the following Hamiltonians \cite{supp}: (1) $H_e(t)=iE(\hat{a}^{\dagger}e^{i\Delta t}-\hat{a}e^{-i\Delta t})$ for an external drive with the intensity $E$ and the detuning $\Delta=\omega_c-\omega_l$ of its frequency $\omega_l$; 
(2) $H_{om}(t)=-g\hat{a}^\dagger\hat{a}(\hat{b}e^{-i\omega_m t}+\hat{b}^\dagger e^{i\omega_m t})$ for the optomechanical coupling; (3) the stochastic Hamiltonian $H_{sr}(t)=i\sqrt{2\kappa }\{\hat{a}^{\dagger }\hat{\xi}
_{c}(t)-\hat{a}\hat{\xi}^{\dagger }
_{c}(t)\}+i\sqrt{2\gamma_m }\{\hat{b}^{\dagger }\hat{\xi}
_{m}(t)-\hat{b}\hat{\xi}^{\dagger }
_{m}(t)\}$ accounting for the cavity (mechanical) damping at the rate $\kappa$ ($\gamma_m$), with the correlations of the quantum noise operators satisfying \cite{rev} 
\begin{eqnarray}
&&\langle \hat{\xi}^\dagger_{c}(t)\hat{\xi}_{c}(t')\rangle=0, ~~\langle \hat{\xi}^\dagger_{m}(t)\hat{\xi}_{m}(t')\rangle=n_{th}\delta(t-t'),
\label{noise}
\end{eqnarray}
where a zero thermal occupation for the cavity reservoir is assumed and $n_{th}$ is the thermal occupation of the mechanical reservoir. Then we take an interaction picture with respect to $H_e(t)$. 
The transformed Hamiltonian $H^{in}(t)=U_0^\dagger(t)(H_{om}+H_{sr})U_0(t)$, where $U_0(t)={\cal T}e^{-i\int_0^t d\tau H_e(\tau)}$, gives the dynamical equations
\begin{eqnarray}
\dot{\hat{a}}&=&-\kappa\hat{a}+g E f(t)e^{-i\omega_mt} \hat{b}+g Ef(t)e^{i\omega_mt} \hat{b}^\dagger\nonumber\\
&+&i\kappa f(t) E+\sqrt{2\kappa}\hat{\xi}_c(t),\\
\dot{\hat{b}}&=&-\gamma_m\hat{b}-g Ef^{\ast}(t)e^{i\omega_mt}\hat{a}+g E f(t)e^{i\omega_mt} \hat{a}^\dagger\nonumber\\
&+&ig|f(t)E|^2e^{i\omega_m t}+\sqrt{2\gamma_m}\hat{\xi}_m(t),
\end{eqnarray}
with $f(t)=(e^{i\Delta t}-1)/\Delta$. A cubic term in $H^{in}(t)$ is neglected in deriving these linear dynamical equations, since we are dealing with a weakly coupled OMS \cite{supp}.

We first look at the mechanical oscillator's thermal equilibrium with the environment. In this initially prepared state with no external drive ($E=0$), the mechanical mode takes the exact form $\hat{b}(t)=e^{-\gamma_m t}\hat{b}+\sqrt{2\gamma_m}\int_0^t d\tau e^{-\gamma_m(t-\tau)}\hat{\xi}_{m}(\tau)$ from Eq. (3). It is the second noise drive term that maintains the invariant phonon number $\langle \hat{b}^\dagger\hat{b}\rangle=n_{th}$ under the thermal equilibrium, while the contribution from the first term lowers with time. The noise actions are thus essential to a quantum OMS.

In a general situation it is clearer to use the equation
\begin{eqnarray}
\frac{d}{dt}\vec{\hat{c}}(t)=\hat{M}(t)\vec{\hat{c}}(t)+\vec{\lambda}(t)+\vec{\hat{\eta}}(t)
\label{d-equation}
\end{eqnarray}
about the complete 
set $\vec{\hat{c}}(t)=(\hat{a}(t),\hat{a}^\dagger(t),\hat{b}(t),\hat{b}^\dagger(t))^T$ of the system modes,
where $\vec{\lambda}=(\lambda_c,\lambda_c^\ast, \lambda_m, \lambda_m^\ast)^T$ with $\lambda_c(t)=i\kappa f(t)E$ and
$\lambda_m(t)=ig|f(t)E|^2e^{i\omega_mt}$, $\vec{\hat{\eta}}=(\hat{\eta}_c, \hat{\eta}_c^\dagger, \hat{\eta}_m,\hat{\eta}_m^\dagger)^T$ with $\hat{\eta}_c(t)=\sqrt{2\kappa}\hat{\xi}_c(t)$ and $\hat{\eta}_m(t)=\sqrt{2\gamma_m}\hat{\xi}_m(t)$,
and the detailed matrix $\hat{M}(t)$ can be found in \cite{supp}.
The general solution to Eq. (\ref{d-equation}) reads
\begin{eqnarray}
\vec{\hat{c}}(t)&=&{\cal T}e^{\int_0^t d\tau \hat{M}(\tau)}\vec{\hat{c}}\nonumber\\
&+&\int_0^t d\tau~ {\cal T}e^{\int_\tau^t dt' \hat{M}(t')}\big(\vec{\lambda}(\tau)+\vec{\hat{\eta}}(\tau)\big),
\label{solution}
\end{eqnarray}
with $\vec{\hat{c}}=(\hat{a},\hat{a}^\dagger,\hat{b},\hat{b}^\dagger)^T$. The time-ordered exponentials appear because we have $[\hat{M}(t),\hat{M}(t')]\neq 0$ for $t\neq t'$. In terms of the notation $d_{ij}(t,\tau)$ for the matrix elements $[{\cal T}e^{\int_\tau^t dt' \hat{M}(t')}]_{ij}$, the evolving thermal phonon number $n_m(t)=\langle \hat{b}^\dagger\hat{b}(t)\rangle-|\langle \hat{b}(t)\rangle|^2$ (the subtraction of the coherent drive $\vec{\lambda}(t)$'s contribution is to be discussed below) consists of two parts,
\begin{eqnarray}
n^{(s)}_m(t)&=&|d_{41}(t,0)|^2+|d_{44}(t,0)|^2n_{th}\nonumber\\
&+&|d_{43}(t,0)|^2(n_{th}+1)
\label{s}
\end{eqnarray}
from taking the expectation value of the first term on the right side of Eq. (\ref{solution}) with respect to the initial cavity vacuum state and mechanical thermal state, and
 \begin{eqnarray}
 n^{(n)}_m(t)&=&2\kappa\int_0^{t} d\tau |d_{41}(t,\tau)|^2+2\gamma_m \int_0^{t} d\tau |d_{44}(t,\tau)|^2n_{th}\nonumber\\
&+& 2\gamma_m \int_0^{t} d\tau |d_{43}(t,\tau)|^2(n_{th}+1)
\label{n}
\end{eqnarray}
from averaging the following noise drive term over the reservoir states by means of Eq. (\ref{noise}).
A meaningful scenario beyond thermal equilibrium is cooling---the noise contribution in Eq. (\ref{n}) finally stabilizes to a thermal phonon number $n_{m,f}$ less than the initial occupation $n_{th}$, while the contribution $n_m^{(s)}(t)$ from the evolved system operators gradually tends to zero with time. 

Going back to the specific terms in Eqs. (2)-(3), one finds that an increased magnitude of the third term on their 
right sides (from a squeezing type coupling) can enhance the phonon numbers in both Eq. (\ref{s}) and Eq. (\ref{n}). 
The cooling action, on the other hand, manifests as the second term of a BS coupling in these equations. 
To let the BS action dominate, one could set the detuning $\Delta$ to be the mechanical frequency $\omega_m$, so that the factor $f(t)e^{-i\omega_mt}$ in Eq. (2) will become $\Gamma_b(t)/\omega_m=(1-e^{-i\omega_m t})/\omega_m$ containing a non-oscillating term. 
Meanwhile, the factor $f(t)e^{i\omega_mt}$ in the squeezing coupling term 
will be $\Gamma_s(t)/\omega_m=(e^{i2\omega_m t}-e^{i\omega_m t})/\omega_m$. A cooling can be performed with a sufficiently large $\omega_m/\kappa$, because in Eq. (5) it suppresses the integrals of the matrix elements carrying the oscillating factor $\Gamma_s(t)$. 

\begin{figure}[t!]
\vspace{-0cm}
\centering
\epsfig{file=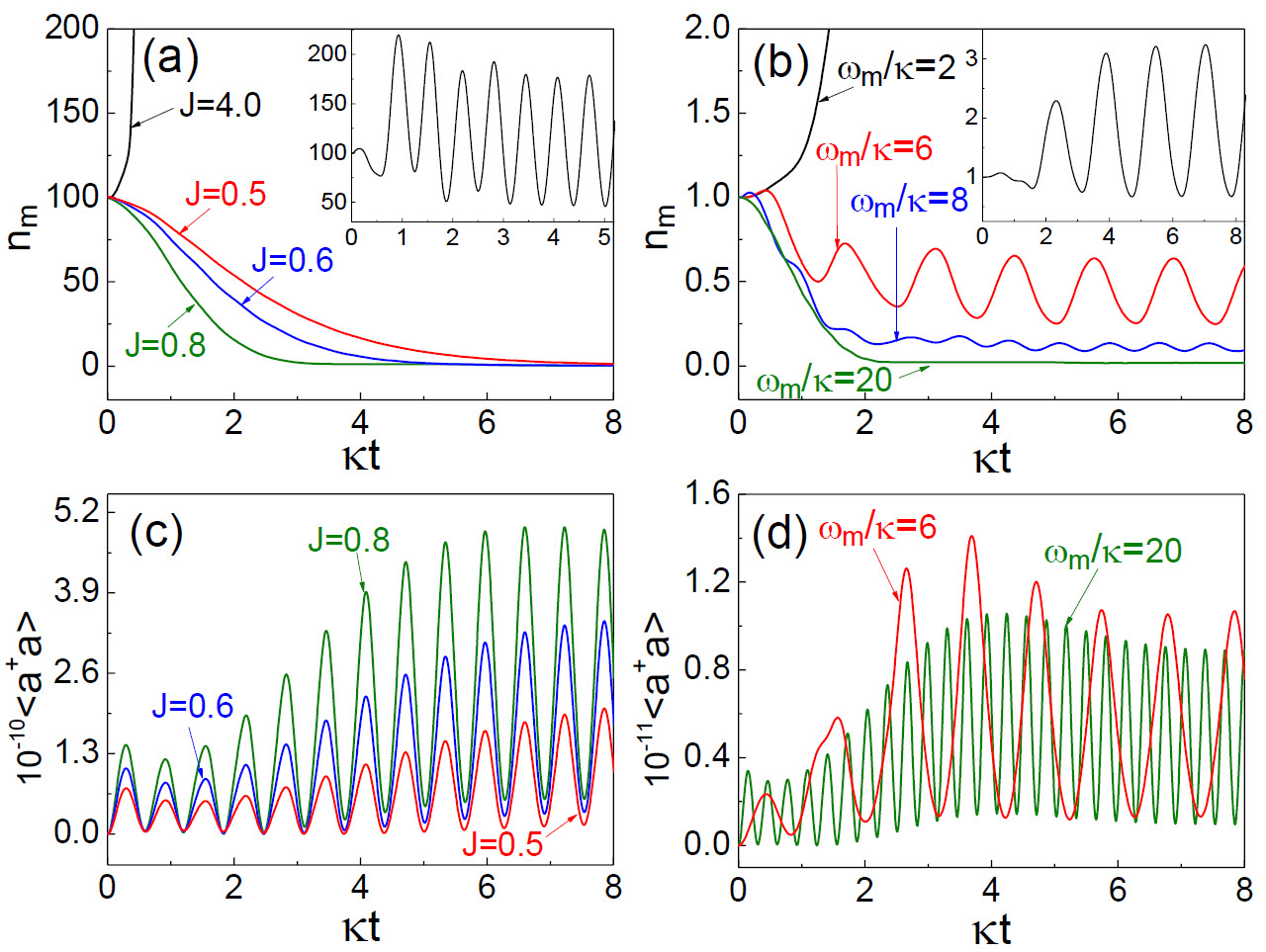,width=1\linewidth,clip=} 
{\vspace{-0.7cm}
\caption{Time evolutions of the thermal phonon number $n_{m}$ and the corresponding cavity photon number $\langle \hat{a}^\dagger\hat{a}\rangle$ 
for the OMSs with $g/\kappa=10^{-5}$, $\gamma_m/\kappa=10^{-3}$, and $\Delta/\omega_m=1$. In (a) and (c) we set $\omega_m/\kappa=10$ and $n_{th}=100$ for the different parameter $J=(g/\omega_m) (E/\kappa)$; in (b) and (d) we take $J=1.0$ and $n_{th}=1$. The insets in (a) and (b) describe the transitional regimes between cooling and heating, about $J=2.5$ and $\omega_m/\kappa=4$, respectively. The different amplitudes in (c) come from the different ratios $E/\omega_m$ for the values of $J$. For clarity only two plots to be stabilized are given in (d).}}
\vspace{-0cm}
\end{figure}

Our dynamical approach directly gives the picture of how the changes of the system parameters will turn the system from heating to cooling; see the time evolutions of the thermal phonon numbers in Figs. 2(a) and 2(b). In the transitional regimes where the cooling (BS) and heating (squeezing) effect compete with each other, the phonon numbers exhibit oscillations with time, as those in the insets of the figures. The corresponding cavity photon numbers in Figs. 2(c) and 2(d) evolve synchronously with the thermal phonon numbers, and the phenomenon is discussed further in \cite{supp}. 

New understanding of optomechanical cooling can be obtained from this dynamical picture, as it brings about another important parameter for cooling in addition to the widely concerned sideband resolution $\omega_m/\kappa$. A relevant phenomenon shown in Fig. 3(a) is that a higher sideband resolution beyond a certain value will be actually worse for cooling if the drive intensity $E$ is fixed. Our predicted thermal phonon number $n_{m,f}$ comes from the noise contributions as the integrals in Eq. (\ref{n}), and is therefore determined by the whole process from $t=0$ to the approximate end time $t=t_s$ when $n_{m}^{(n)}(t)$ begins to be stable and $n_{m}^{(s)}(t_s)\rightarrow 0$ (the associated dynamical behaviors at $t_s$ asymptotically approach those at $t=\infty$). 
A slower process will accumulate a higher noise contribution $n_{m,f}$, which can be the time average of $n_{m}^{(n)}(t)$ ($t>t_s$) for the finally oscillating ones in Fig. 2(b). At the BS coupling resonance $\Delta=\omega_m$, the coefficient of the BS (squeezing) coupling term in Eqs. (2)-(3) takes the form  $J \Gamma_{b(s)}(\kappa t)$. The parameter defined as $J=(g/\omega_m)\cdot (E/\kappa)$ decides how fast the cooling of a specific OMS will be; see the illustration in Fig. 2(a). Were there no quantum noises, an uncoupled system with $J=0$ would take the longest time scale $1/\gamma_m$ ($\gamma_m\ll \kappa$) to reach its stability; cf. the solution $\hat{b}(t)=e^{-\gamma_m t}\hat{b}$ to Eq. (3) when $E, \hat{\xi}_m(t)=0$. With the drive intensity $E$ fixed as in Fig. 3(a), a lower sideband resolution $\omega_m/\kappa$ corresponding to a higher $J$ makes the cooling faster. It is also shown in Fig. 2(b) that, once the parameter $J$ is fixed, the OMSs with different sideband resolution $\omega_m/\kappa$ will evolve to the stabilized phase almost together. 

\begin{figure}[t!]
\vspace{-0cm}
\centering
\epsfig{file=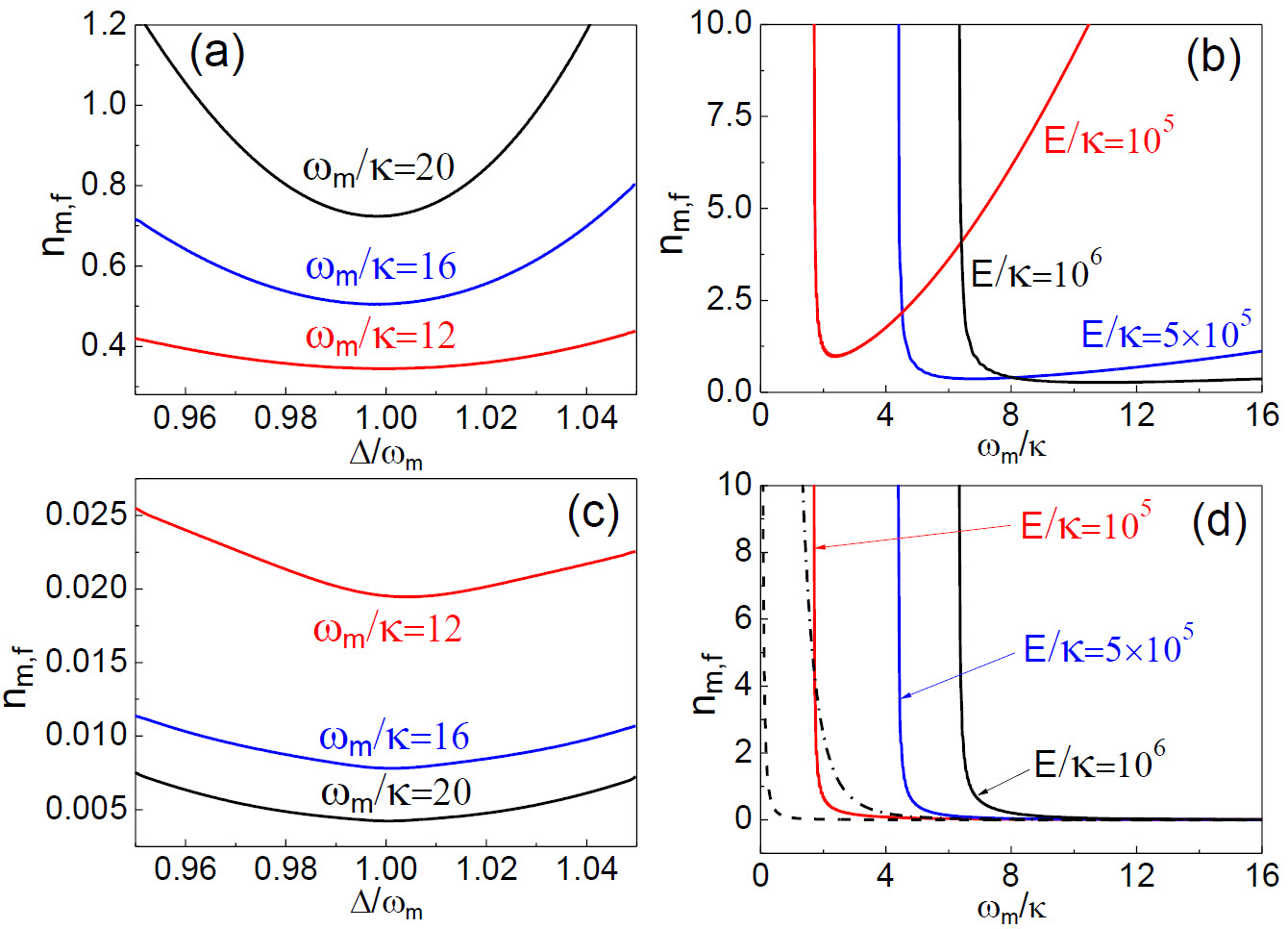,width=1\linewidth,clip=} 
{\vspace{-0.6cm}
\caption{Stabilized thermal phonon numbers achievable for the OMSs with $g/\kappa=10^{-5}$ and $\gamma_m/\kappa=10^{-3}$. 
In (a) and (b) the initial thermal phonon number $n_{th}$ is $100$, and in (c) and (d) the systems are assumed to be with $n_{th}=0$. The drive intensity is set at $E=8\times 10^5\kappa$ in (a) and (c). The detuning $\Delta$ equals to $\omega_m$ in (b) and (d). In (d) the dashed curve is $n_{m,f}=\kappa^2/(4\omega_m^2)$, corresponding to $\kappa^2/(16\omega_m^2)$ in \cite{c1,c2} due to a different expression for the cavity damping term. The dot-dashed curve obtained with $E/\kappa=10^6$ 
is the previous prediction for the strong coupling regime \cite{c3}.}}
\vspace{-0cm}
\end{figure}

Apart from speeding up the cooling of an OMS, increasing $J$ will make the coexisting squeezing (heating) effect stronger. These two tendencies strike a balance somewhere in the parameter space, so that the best cooling under a fixed drive intensity $E$ takes place at an optimum $\omega_m/\kappa$ [see Fig. 3(b)]. The latter tendency will dominate when the parameter $\omega_m/\kappa$ is continuously lowered, 
which will increase $n_{m,f}$ significantly on the left of the optimum $\omega_m/\kappa$. On the other hand, the suppression of the factor $\Gamma_s(t)$ by faster oscillations due to larger $\omega_m/\kappa$ will diminish the squeezing effect, leading to the tendency of the stabilized $n_m$ in Fig. 2(b). Among the three different terms in Eq. (7), the first one independent of $n_{th}$ monotonically decreases with increased $\omega_m/\kappa$, as illustrated in Figs. 3(c) and 3(d). The experimental investigation of the effect related to this contribution was recently reported in \cite{ex12}. Such pure cavity noise contribution decided by the squeezing coupling intensity $J\Gamma_s(\kappa t)$ increases significantly with the drive intensity $E$, which distinguishes our results fundamentally from the previous predictions; see the comparisons in Fig. 3(d) and Fig. S-4(b) in \cite{supp}. 

Another important issue is how to reach the best cooling from the initial thermal equilibrium of an OMS.
To answer the question, we first examine the limit of $\omega_m/\kappa\rightarrow \infty$. In this limit the 
stabilized thermal phonon number given by $\Delta=\omega_m$ can be found analytically from Eqs. (2)-(3) as
\begin{eqnarray}
n_{m,f}
&=& \Gamma_m n_{th}\frac{1}{|\eta_{+}-\eta_{-}|^2}\nonumber\\
&\times & \big\{4\mbox{Re} \{\frac{\eta_{+}^\ast \eta_{-}}{(\lambda_+^\ast+\lambda_-)}\}-\frac{|\eta_{+}|^2}{\mbox{Re}\lambda_{-}}-\frac{|\eta_{-}|^2}{\mbox{Re}\lambda_{+}}\big\},
\label{l}
\end{eqnarray}
where $\lambda_{\pm}=\frac{1}{2}(-1-\Gamma_m\pm \sqrt{(1-\Gamma_m)^2-4J^2})$ and $\eta_{\pm}=-1+\Gamma_m\pm \sqrt{(1-\Gamma_m)^2-4J^2}$ with $\Gamma_m=\gamma_m/\kappa$. It is solely from the contribution of the second term 
in Eq. (\ref{n}), since the squeezing action with its intensity $J\Gamma_s(\kappa t)$ has been completely averaged out by 
an infinite $\omega_m/\kappa$. Analogous to a first-order phase transition, this limiting value exhibits a discontinuous jump across the point $J=1/2(1-\Gamma_m)$ (in Fig. 4 it is around $J=0.5$), which separates the weak coupling regime from the strong coupling one that has been experimentally observed \cite{ex8, ex10, strong}. In the strong coupling regime, the limiting value becomes the constant $n_{m,f}=\Gamma_m n_{th}$. Inside the weak coupling regime the stabilized phonon numbers for different $\omega_m/\kappa$ can be close to the limiting value and drop quickly with increased $J$ (see the inset in Fig. 4). With the parameters of the experimental setup in \cite{ex9} (equivalent to $J\approx 0.16$), for example, an ideal cooling process would not reach a phonon number lower than the limit of $n_{m,f}\approx 0.26$. 

\begin{figure}[t!]
\vspace{0cm}
\centering
\epsfig{file=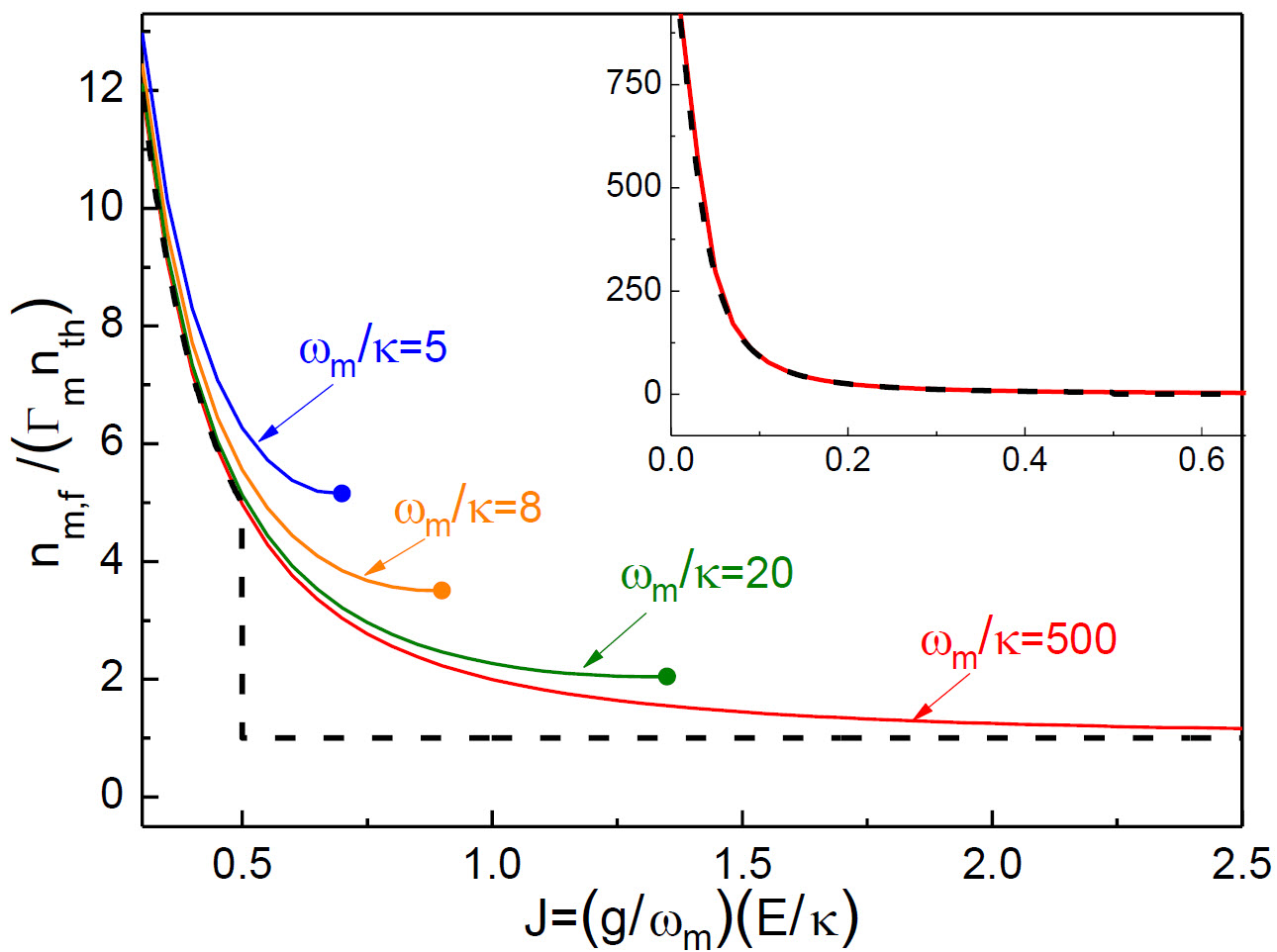,width=0.81\linewidth,clip=} 
{\vspace{-0cm}
\caption{System parameter relations that determine the optimum cooling for a general OMS under the BS coupling resonance $\Delta=\omega_m$ and with a fixed ratio $\Gamma_m=\gamma_m/\kappa=10^{-3}$. The dashed curve represents the cooling limit obtained with $\omega_m/\kappa\rightarrow \infty$. The ending dot for each curve is the point of the optimum cooling for a fixed $\omega_m/\kappa$. These curves asymptotically tend to the cooling limit (the dashed one) with increased $\omega_m/\kappa$. The tendency of the lower $J$ is shown in the inset, having all curves stuck together as viewed with the range of the obtained result.}}
\vspace{-0.3cm}
\end{figure}

The parametric conditions for achieving good cooling manifest more clearly with Fig. 4, in which the limiting value in 
Eq. (\ref{l}) gives the boundary any cooling process cannot surpass. For a setup with the built-in 
sideband resolution $\omega_m/\kappa$, increasing the drive power can make its cooling better and faster, moving the achieved thermal phonon number along a curve of fixed $\omega_m/\kappa$ to larger $J$, though the enhanced power may change the phonon number less obviously for a system with $\omega_m/\kappa\gg 1$ (in the strong coupling regime its phonon curve asymptotically approaches the limit line). The higher the parameter $\omega_m/\kappa$ is, the stronger the drive can be applied to achieve a better cooling; 
this explains the phonon number tendency on the right of the optimum points in Fig. 3(b). However, the cooling cannot be improved further if $J$ arrives at a turning point (the ending dots of the phonon curves) where the coexisting squeezing effect begins to be significant. After crossing the point, the thermal phonon number will increase from the minimum value there, and its evolution will continue to become oscillating and then growing with time as in Fig. 2(a). Such a reference point indicates that the cooling processes in Fig. 2(b) are not the optimum ones---the parameter $J=1.0$ used there has been on the right side of the turning point for $\omega_m/\kappa=8$. 

The evolving Gaussian states of the weakly coupled OMSs, which follow the linear dynamical equations, can be depicted with their Wigner functions. A cooling process starts from a mechanical oscillator's thermal state with its Wigner function being 
$W(q_m, p_m)=[1/\pi(1+2n_{th})]\exp\{-[1/(1+2n_{th})](q_m^2+p_m^2)\}$. After turning on a drive, the OMS will evolve under optomechanical coupling to two-mode Gaussian states, whose Wigner functions can be found by numerically calculating the correlation matrix of the two system modes. As illustrated in Fig. 1(b), a perfect ``ground-state cooling" to $n_{m,f}=0$ evolves the mechanical mode's Wigner function to $W(q_m, p_m)=\frac{1}{\pi}e^{-(q_m-q_m^0)^2-(p_m-p_m^0)^2}$ of a coherent state, which differs from that of a vacuum state only by the displacements $q_m^0(t)$, $p_m^0(t)$ determined by the coherent drive term $\vec{\lambda}(t)$ in Eq. (4). Without impairing the purity \cite{purity} of the target coherent state, the contribution from the coherent drives should be excluded from the thermal phonon number $n_{m}(t)$ to be reduced from the occupation $n_{th}$ of an initial thermal state. 
 
In summary, we have developed a quantum dynamical approach to optomechanical cooling. The motivation for the development is to reflect the fact that such cooling is a process for a mechanical oscillator to evolve from its initial thermal state 
to another state with higher purity, which takes time. Cooling starts after turning on a red-detuned drive that realizes a BS type coupling between the fast damping cavity mode and the slowly decaying mechanical mode, so that the oscillator's thermal excitation being converted to cavity photons could be totally eliminated after a time controlled by the parameter $J$. Only with this scenario giving the phonon number in Eq. (\ref{s}), the mechanical oscillator would be cooled down to a vacuum state modified by its inevitable motion under radiation pressure to a coherent state. The simultaneous quantum noise actions, however, cause the evolution to deviate from going to such a pure quantum state and add to the thermal phonon number in Eq. (\ref{n}). The thermal occupation left in the end is decided by how soon the coupled cavity and mechanical modes evolve together to a dynamical stability, after which the system cannot be cooled down further. Like the fully quantum mechanical treatment of OMS in \cite{law} and recent studies of the quantum dynamical features of other physical systems (see, e.g. \cite{dd1,dd2}), the properties in a cooling process illustrated here are for a genuine quantum OMS rather than the quantum fluctuations around the trajectories of a classical one; this approach can also be used for blue-detuned drives \cite{s-laser}. This quantum dynamical picture of cooling applies to an OMS truly approaching its macroscopic quantum states.

We acknowledge the funding supports from NSFC (Grant No. 11574093 and Grant No. 61435007). L. Y. is sponsored by the Fundamental Research Funds for the Central Universities (No. HEUCFJ170402). This research is also supported by the Arkansas High Performance Computing Center and the Arkansas Economic Development Commission.

B. H. and L. Y. contributed equally to this work.

\newpage

\begin{widetext}

\section*{{\large Supplementary Material for ``Radiation Pressure Cooling as a Quantum Dynamical Process"}}

\renewcommand{\thefigure}{S-\arabic{figure}}
\setcounter{figure}{0}

\subsection{I. DYNAMICAL EQUATIONS OF OPTOMECHANICAL SYSTEMS IN A ROTATING FRAME}
\renewcommand{\theequation}{S-I-\arabic{equation}}
\setcounter{equation}{0}

\renewcommand{\theequation}{S-I-\arabic{equation}}
\setcounter{equation}{0}

A quantum optomechanical system (OMS) is modeled by the coupled cavity mode $\hat{a}$ with the frequency $\omega_c$ and mechanical mode $\hat{b}$ with the frequency $\omega_m$, having the self-oscillation Hamiltonians $\omega_c\hat{a}^\dagger\hat{a}$ and $\omega_m\hat{b}^\dagger\hat{b}$ ($\hbar=1$) for the respective mode. Initially, when there is no external drive field, the mechanical oscillator is in a thermal equilibrium with the environment 
at a temperature corresponding to the occupation $n_{th}$. 
An external drive with the intensity $E$ and frequency $\omega_l$, as seen from the Hamiltonian $iE(\hat{a}^{\dagger}e^{-i\omega_l t}-\hat{a}e^{i\omega_l t})$, will build up the cavity field. Under the radiation pressure of the cavity field, the size of the cavity will be changed from $L$ to $L+\hat{x}_m$, where $\hat{x}_m=\sqrt{\frac{1}{2m\omega_m}}(\hat{b}+\hat{b}^\dagger)$ 
is the displacement of the mechanical oscillator with the effective mass $m$ 
(the notation of $\hbar=1$ is adopted), and is treated as a q-number 
for a quantum OMS. The consequentially modified cavity oscillation term $n\pi c/(L+\hat{x}_m)\hat{a}^\dagger\hat{a}\approx \omega_c\hat{a}^\dagger\hat{a}-(\omega_c/L)\hat{x}_m\hat{a}^\dagger\hat{a}$ gives the optomechanical coupling Hamiltonian as the second term written 
as $-g\hat{a}^\dagger\hat{a}(\hat{b}+\hat{b}^\dagger)$, where $g=\sqrt{\frac{1}{2m\omega_m}}\frac{\omega_c}{L}$.

In addition to the above-mentioned elements, the OMS is an open quantum system coupled to the environment known as the cavity and mechanical reservoirs, which are modeled as ensembles of oscillators with the Hamiltonian 
\begin{eqnarray}
H_r=\int_0^\infty d\omega_1\omega_1\hat{\xi}^\dagger_c\hat{\xi}_c(\omega_1)+\int_0^\infty d\omega_2\omega_2\hat{\xi}^\dagger_m\hat{\xi}_m(\omega_2)
\end{eqnarray}
where $[\hat{\xi}_{c,m}(\omega),\hat{\xi}^\dagger_{c,m}(\omega')]=\delta(\omega-\omega')$.
The Hamiltonian for the coupling between the system and the reservoirs takes the general form
\begin{eqnarray}
H_{s-r}=i\int d\omega_1 \kappa(\omega_1)(\hat{a}-\hat{a}^\dagger)\{\hat{\xi}^\dagger_c(\omega_1)+\hat{\xi}_c(\omega_1)\}+i\int d\omega_2 \gamma_m(\omega_2)(\hat{b}-\hat{b}^\dagger)\{\hat{\xi}^\dagger_m(\omega_2)+\hat{\xi}_m(\omega_2)\}.
\end{eqnarray}
In the rotating frame with respect to the cavity frequency $\omega_c$ and the mechanical frequency $\omega_m$, as well as to the frequencies of 
the reservoir modes, the system-reservoir coupling Hamiltonian will become
\begin{eqnarray}
H^{(i)}_{s-r}&=&i\int d\omega_1 \kappa(\omega_1)(\hat{a}e^{-i\omega_c t}-\hat{a}^\dagger e^{i\omega_c t})\{\hat{\xi}^\dagger_c(\omega_1)e^{i\omega_1 t}+\hat{\xi}_c(\omega_1)e^{-i\omega_1 t}\}\nonumber\\
&+&i\int d\omega_2 \gamma_m(\omega_2)(\hat{b}e^{-i\omega_m t}-\hat{b}^\dagger e^{i\omega_m t})\{\hat{\xi}^\dagger_m(\omega_2)e^{i\omega_2 t}+\hat{\xi}_m
(\omega_2)e^{-i\omega_2 t}\},
\label{couple}
\end{eqnarray}
as the result of the unitary transformation with $U_I(t)=\exp\{-i\int_0^t d\tau (H_r+\omega_c\hat{a}^\dagger\hat{a}+\omega_m\hat{b}^\dagger\hat{b})\}$. 
Similar to the practices in \cite{supp1}, we make a rotating-wave approximation to neglect the rapidly oscillating terms in Eq. (\ref{couple}), together with an approximation of smooth system-reservoir couplings to reduce the associated 
coupling intensities to constants, i.e, $\kappa(\omega)\rightarrow \sqrt{\kappa/\pi}$ and $\gamma_m(\omega)\rightarrow \sqrt{\gamma_m/ \pi}$. Then the system-reservoir coupling will be simplified to
\begin{eqnarray}
H_{sr}(t)=i\sqrt{2\kappa }\{\hat{a}^{\dagger }\hat{\xi}_{c}(t)-\hat{a}\hat{\xi}^{\dagger}_{c}(t)\}+i\sqrt{2\gamma_m }\{\hat{b}^{\dagger }
\hat{\xi}_{m}(t)-\hat{b}\hat{\xi}^{\dagger }_{m}(t)\},
\label{couple1}
\end{eqnarray}
where $\hat{\xi}_{c,m}(t)=\frac{1}{\sqrt{2\pi}}\int d\omega \hat{\xi}_{c,m}(\omega)e^{-i(\omega-\omega_{c,m})t}$.
For the relatively slow interaction processes described by Eq. (\ref{couple1}), the quantum noise represented 
by $\hat{\xi}_{c,m}(t)$ can be approximated as a white one 
satisfying $[\hat{\xi}_{c,m}(t),\hat{\xi}^\dagger_{c,m}(t')]=\delta(t-t')$ \cite{supp1}. 

Meanwhile, under the above-mentioned rotation by $U_I(t)$, the external drive Hamiltonian and the optomechanical coupling Hamiltonian will become $H_e(t)=iE(\hat{a}^{\dagger}e^{i\Delta t}-\hat{a}e^{-i\Delta t})$, where $\Delta=\omega_c-\omega_l$, and $H_{om}(t)=-g\hat{a}^\dagger\hat{a}(\hat{b}e^{-i\omega_m t}+\hat{b}^\dagger e^{i\omega_m t})$, respectively.
The action $U(t)={\cal T}\exp\{-i\int_0^t d\tau H(\tau)\}$ of the total Hamiltonian $H(t)=H_e(t)+H_{om}(t)+H_{sr}(t)$, which is called the formal solution to Quantum Stochastic Schr\"{o}dinger equations in \cite{supp1}, will lead to the following nonlinear dynamical equations of the system modes:
\begin{eqnarray}
\dot{\hat{a}}&=&-\kappa\hat{a}+ig(\hat{b}e^{-i\omega_m t}+\hat{b}^\dagger e^{i\omega_m t})\hat{a}+E e^{i\Delta t}+\sqrt{2\kappa}\hat{\xi}_c(t),\nonumber\\
\dot{\hat{b}}&=&-\gamma_m\hat{b}+ige^{i\omega_m t}\hat{a}^\dagger\hat{a}+\sqrt{2\gamma_m}\hat{\xi}_m(t).
\label{e1}
\end{eqnarray}
Among the literature, the dynamics of an OMS is also described in another rotating frame only with respect to the drive frequency $\omega_l$ 
(see, e.g. \cite{rev}). Then its dynamical equations read
\begin{eqnarray}
\dot{\hat{a}}&=&-\kappa\hat{a}-i\Delta \hat{a}+ig(\hat{b}+\hat{b}^\dagger)\hat{a}+E +\sqrt{2\kappa}\hat{\xi}_c(t),\nonumber\\
\dot{\hat{b}}&=&-\gamma_m\hat{b}-i\omega_m\hat{b}+ig\hat{a}^\dagger\hat{a}+\sqrt{2\gamma_m}\hat{\xi}_m(t),
\label{e2}
\end{eqnarray}
which directly uses the system-reservoir coupling Hamiltonian in Eq. (\ref{couple1}).

\subsection{II. LINEARIZED DYNAMICS OF WEAKLY COUPLED QUANTUM OPTOMECHANICAL SYSTEMS}
\renewcommand{\theequation}{S-II-\arabic{equation}}
\setcounter{equation}{0}
For the total Hamiltonian $H(t)=H_e(t)+H_{om}(t)+H_{sr}(t)$ in our used rotating frame, 
we take another interaction picture with respect to the 
part of $H_e(t)$, to have the cavity mode $\hat{a}$ displaced by the unitary operation $U_0(t)={\cal T}\exp\{-i\int_0^t d\tau H_e(\tau)\}$ 
to $\hat{a}-i\frac{e^{i\Delta t}-1}{\Delta}E$. Then the remaining part of the Hamiltonian $H(t)$ will be transformed to
\begin{eqnarray}
H^{in}(t)&=& U_0^\dagger(t)\big(H_{om}(t)+H_{sr}(t)\big)U_0(t)\nonumber\\
&=&\underbrace{ig E\frac{e^{i\Delta t}-1}{\Delta}e^{-i\omega_mt} \hat{a}^\dagger \hat{b}-ig  E \frac{e^{-i\Delta t}-1}{\Delta}e^{i\omega_mt} \hat{a} \hat{b}^\dagger
+ig E\frac{e^{i\Delta t}-1}{\Delta}e^{i\omega_mt} \hat{a}^\dagger \hat{b}^\dagger-ig E \frac{e^{-i\Delta t}-1}{\Delta}e^{-i\omega_mt} \hat{a} \hat{b}}_{H_{couple}(t)}\nonumber\\
&-&g\underbrace{\big|\frac{e^{i\Delta t}-1}{\Delta}E\big|^2(e^{-i\omega_mt}\hat{b}+e^{i\omega_m t}\hat{b}^\dagger)}_{H_{displace}(t)}
+i\sqrt{2\kappa}\{(\hat{a}^\dagger +i\frac{e^{-i\Delta t}-1}{\Delta}E)\hat{\xi}_p(t)
-(\hat{a}-i\frac{e^{i\Delta t}-1}{\Delta}E)\hat{\xi}_p^\dagger(t)\}\nonumber\\
&+&i\sqrt{2\gamma_m }\{\hat{b}^{\dagger}\hat{\xi}_{m}(t)-\hat{b}\hat{\xi}_{m}^\dagger (t)\}
-\underbrace{g\hat{a}^\dagger\hat{a}(e^{-i\omega_mt}\hat{b}+e^{i\omega_m t}\hat{b}^\dagger)}_{H_{nl}(t)}.
\label{Hi}
\end{eqnarray}
Among this transformed Hamiltonian $H^{in}(t)$, the effect of the cubic term $H_{nl}(t)$ is insignificant compared with the others. Because it is non-commutative with the rest of the Hamiltonian, its action will modify the mode operators in the rest part of $H^{in}(t)$ by the unitary action \cite{s-laser}
\begin{eqnarray}
U_{nl}(t,\tau)&=&{\cal T}e^{i\int_\tau^t dt' g\hat{a}^\dagger\hat{a}(e^{-i\omega_m t'}\hat{b}+e^{i\omega_m t'}
\hat{b}^\dagger)}\nonumber\\
&=& e^{i\int_\tau^t dt' g\hat{a}^\dagger\hat{a}\hat{b}e^{-i\omega_m t'}}e^{i\int_\tau^t dt' g\hat{a}^\dagger\hat{a}\hat{b}^\dagger e^{i\omega_m t'}}
e^{i\int_\tau^t dt'\frac{g^2}{\omega_m}(1-e^{i\omega_m(t'-\tau)})(\hat{a}^\dagger\hat{a})^2},
\label{NL}
\end{eqnarray}
as it can be seen from the expectation value of any system operator $\hat{O}$:
\begin{eqnarray}
\langle \hat{O}(t)\rangle&=&\mbox{Tr}_s\{\hat{O}\rho(t)\}=\mbox{Tr}_s\{\hat{O}\mbox{Tr}_r(U(t)\rho(0)\rho_r U^\dagger(t))\}\nonumber\\
&=&\mbox{Tr}_{s,r}\{{\cal T}e^{i\int_0^t d\tau U_{nl}(t,\tau)\{H^{in}(\tau)+H_{nl}(\tau)\}U^\dagger_{nl}(t,\tau)}U^\dagger_0(t)\hat{O}U_0(t){\cal T}e^{-i\int_0^t d\tau U_{nl}(t,\tau)\{H^{in}(\tau)+H_{nl}(\tau)\}U^\dagger_{nl}(t,\tau)}\nonumber\\
&\times & (U_{nl}(t,0)\rho(0)\rho_r U^\dagger_{nl}(t,0))\}\nonumber\\
&\approx &\mbox{Tr}_{s,r}\{{\cal T}e^{i\int_0^t d\tau \{H^{in}(\tau)+H_{nl}(\tau)\}}U^\dagger_0(t)\hat{O}U_0(t){\cal T}e^{-i\int_0^t d\tau \{H^{in}(\tau)+H_{nl}(\tau)\}}\rho(0)\rho_r\},
\label{expect}
\end{eqnarray}
where $U(t)={\cal T}\exp\{-i\int_0^t d\tau H(\tau)\}$ is the action including that of the stochastic part $H_{sr}(t)$ \cite{supp1}, and $\rho_r$ represents the total quantum state of the reservoirs. Because a concerned dynamical evolution starts from a vacuum state $|0\rangle_c$ of the cavity and a thermal state $\rho_m(0)=\sum_{n=0}^{\infty} \frac{n_{th}^n}{(1+n_{th})^{n+1}}\left\vert n\right\rangle_m\left\langle n\right\vert$ of the mechanical oscillator 
($\rho(0)$ is their product and $n_{th}$ is the initial thermal occupation), the action $U_{nl}(t,0)\rho(0)U^\dagger_{nl}(t,0)$ keeps this initial state invariant on the third line of Eq. (\ref{expect}), as a consequence of the relation $U_{nl}(t,0)|0\rangle_c=|0\rangle_c$. 
Note that, in Eq. (\ref{expect}), the operators $\hat{c}=\hat{a},\hat{b}$ in $H^{in}(\tau)+H_{nl}(\tau)$ are transformed by the action in Eq. (\ref{NL}) to $U_{nl}(t,\tau)\hat{c}U^\dagger_{nl}(t,\tau)$, unlike the general 
form $U^{\dagger}(t)\hat{o}U(t)$ for a unitary operation on an operator $\hat{o}$. 
According to Eq. (\ref{NL}), the modified operators differ from the original ones $\hat{c}$ by the corrections in the orders of $g/\omega_m\ll 1$, so these corrections can be well neglected for a weakly coupled OMS. This step noted with an approximation sign on the last line of Eq. (\ref{expect}) is the only theoretical approximation made in our approach, in addition to those commonly used ones to obtain the system-reservoir coupling Hamiltonian in Eq. (\ref{couple1}). 
The neglecting of the cubic term $H_{nl}(t)$ has a physically different meaning from that of omitting the higher-order fluctuation terms in the previously used linearization around the steady states of the averaged Eq. (\ref{e2}), because it is not necessary for our completely quantum treatment to require that the cavity mean field should overwhelm the corresponding cavity field fluctuation.

By using the associated Ito's rules for the stochastic part \cite{supp1}, one will obtain the following dynamical equation
\begin{eqnarray}
\frac{d}{dt}\left( 
\begin{array}{c}
\hat{a}\\ 
\hat{a}^\dagger\\
\hat{b}\\
\hat{b}^\dagger
\end{array}
\right) 
&=&
\underbrace{\left( 
\begin{array}{cccc}
-\kappa & 0 & gE f(t)e^{-i\omega_m t} & gE f(t)e^{i\omega_m t}\\ 
0 &-\kappa & gE f^\ast(t)e^{-i\omega_m t}& gE f^\ast(t)e^{i\omega_m t}\\
-gE f^\ast (t)e^{i\omega_m t}&gE f(t)e^{i\omega_m t}&-\gamma_m & 0\\
gE f^\ast(t)e^{-i\omega_m t} &-gE f(t)e^{-i\omega_m t} &0 &-\gamma_m\\
\end{array}
\right)}_{\hat{M}(t)}\left( 
\begin{array}{c}
\hat{a}\\ 
\hat{a}^\dagger\\
\hat{b}\\
\hat{b}^\dagger
\end{array}
\right) \nonumber\\
 &+&
\underbrace{\left( \begin{array}{c}
i\kappa f(t)E \\ 
-i\kappa f^\ast(t)E\\
ig|f(t)E|^2e^{i\omega_mt}\\
-ig|f(t)E|^2e^{-i\omega_mt}\\
\end{array}
\right)}_{\vec{\lambda}(t)}+\underbrace{\left( \begin{array}{c}
\sqrt{2\kappa}\hat{\xi}_c(t) \\ 
\sqrt{2\kappa}\hat{\xi}^\dagger_c(t)\\
\sqrt{2\gamma_m}\hat{\xi}_m(t)\\
\sqrt{2\gamma_m}\hat{\xi}^\dagger_m(t)\\
\end{array}
\right)}_{\vec{\hat{\eta}}(t)}
\label{dynamics}
\end{eqnarray}
with the effective Hamiltonian $H^{in}(t)+H_{nl}(t)$. This extended form of Eqs. (2)-(3) in the main text 
contains the factor $f(t)=\frac{e^{i\Delta t}-1}{\Delta}$ in the dynamical matrix $\hat{M}(t)$ and the coherent drive term $\vec{\hat{\lambda}}(t)$. The numerical solution to this differential equation can be found by calculating the time-ordered exponentials (in Eq. (5) of the main text) as the products of the matrices $(I+\hat{M}(t_i)\delta t)$ at the different $t_i$ \cite{s-laser}, where $\delta t$ is a small iteration step. The corresponding quantum master equation for the reduced system state $\rho(t)$ reads
\begin{eqnarray}
\dot{\rho}(t)&=&-i[H_{couple}(t)-H_{displace}(t),\rho(t)]-\kappa\big(\hat{o}_a^\dagger\hat{o}_a\rho(t)+\rho(t)\hat{o}_a^\dagger\hat{o}_a-2\hat{o}_a\rho(t)\hat{o}_a^\dagger\big)\nonumber\\
&-&\gamma_m(n_{th}+1)\big(\hat{b}^\dagger\hat{b}\rho(t)+\rho(t)\hat{b}^\dagger\hat{b}-2\hat{b}\rho(t)\hat{b}^\dagger\big)-\gamma_m n_{th}\big(\hat{b}\hat{b}^\dagger\rho(t)+\rho(t)\hat{b}\hat{b}^\dagger-2\hat{b}^\dagger\rho(t)\hat{b}\big)
\end{eqnarray}
with the notations in Eq. (\ref{Hi}), where $\hat{o}_a=\hat{a}-i\frac{e^{i\Delta t}-1}{\Delta}E$.

As a comparison, in the previous approach of linearization with the decomposition $\hat{a}=\alpha+\delta \hat{a}$ or the displacement $\hat{a}\rightarrow  \hat{a}+\alpha$ with a steady-state value $\alpha$ of the cavity mode $\hat{a}$, the linearized Hamiltonian excluding the coherent and noise drive terms takes the form \cite{supp2}
\begin{eqnarray}
H_{linear}=G(\delta \hat{a} e^{-i\Delta t}+\delta \hat{a}^\dagger e^{i\Delta t})(\hat{b}e^{-i\omega_m t}+\hat{b}^\dagger e^{i\omega_m t}),
\label{lnh}
\end{eqnarray}
with $|G|=g|\alpha|$ being a constant. A variation form of this linearized Hamiltonian has been used to study real-time cooling processes \cite{c6}. In reality, however, the constant optomechanical coupling intensity $G$, which equals to $g$ multiplied by the steady value $\alpha$ appearing after a certain period of time, can not exist before that time, for instance, shortly after turning on the cooling laser beam at $t=0$. The cavity field will certainly take time to evolve to stable value. So it is more appropriate to use an evolving field $\alpha(t)$, which develops from zero to stable value throughout a cooling process, in that linearization approach. 
Our quantum dynamical approach totally dispenses with such time-dependent classical mean values, which should be found by solving the nonlinear classical dynamical equations [those obtained by taking the averages of Eq. (\ref{e1}) or Eq. (\ref{e2})]. The linearization of the system dynamics is realized by a unitary transformation $\hat{a}\rightarrow \hat{a}-i\frac{e^{i\Delta t}-1}{\Delta}E$ of the cavity mode, which is completely inequivalent to the previous linearization by the shift $\hat{a}\rightarrow  \hat{a}+\alpha$ with a time-independent classical steady-state value $\alpha$. 

\subsection{III. FURTHER DISCUSSION ON THE EFFECTS OF THE COHERENT DRIVES}
\renewcommand{\theequation}{S-III-\arabic{equation}}
\setcounter{equation}{0}
A majority of physical effects in the concerned cooling processes come from the coherent drive term $\vec{\lambda}(t)$ and the noise drive term $\vec{\hat{\eta}}(t)$ in Eq. (\ref{dynamics}). Here we discuss more of their properties that are complementary to those described in the main text. 
We first look at the former, which has been less explored by the previous studies. In terms of the five dimensionless parameters $G_m=g/\kappa, {\cal E}=E/\kappa, s_m=\omega_m/\kappa, Q=\omega_m/\gamma_m, \delta=\Delta/\omega_m$, as well as the dimensionless time $\tau$ going from $0$ 
to $\kappa t$, the evolving cavity photon number 
$$\langle \hat{a}^\dagger\hat{a}(t)\rangle=\langle U^\dagger(t)\hat{a}^\dagger\hat{a} U(t)\rangle=\langle {\cal T}e^{i\int_0^t d\tau \{H^{in}(\tau)+H_{nl}(\tau)\}}U^\dagger_0(t)\hat{a}^\dagger\hat{a}U_0(t){\cal T}e^{-i\int_0^t d\tau \{H^{in}(\tau)+H_{nl}(\tau)\}}\rangle $$
is given as the sum of the following three parts:
(1) the system operator contribution 
\begin{eqnarray}
\langle \hat{a}^\dagger\hat{a}(t)\rangle_s=d_{21}(\kappa t,0)d_{12}(\kappa t,0)+d_{23}(\kappa t,0)d_{14}(\kappa t,0)(n_{th}+1)+d_{24}(\kappa t,0)d_{13}(\kappa t,0)n_{th};
\label{c1}
\end{eqnarray}
(2) the noise drive contribution 
\begin{eqnarray}
\langle \hat{a}^\dagger\hat{a}(t)\rangle_n&=&2\int_0^{\kappa t} d\tau d_{21}(\kappa t,\tau)d_{12}(\kappa t,\tau)+\frac{2s_m}{Q}\int_0^{\kappa t} d\tau d_{23}(\kappa t,\tau)d_{14}(\kappa t,\tau)(n_{th}+1)\nonumber\\
&+&\frac{2s_m}{Q}\int_0^{\kappa t} d\tau d_{24}(\kappa t,\tau)d_{13}(\kappa t,\tau)n_{th};
\label{c2}
\end{eqnarray}
(3) the contribution 
\begin{eqnarray}
\langle \hat{a}^\dagger\hat{a}(t)\rangle_c &=&\big |{\cal E}\int_0^{\kappa t} d\tau e^{i\delta s_m  \tau}+i \frac{{\cal E}}{s_m \delta}\int_0^{\kappa t} d\tau d_{11}(\kappa t,\tau)F(\tau)-i\frac{{\cal E}}{s_m \delta}\int_0^{\kappa t} d\tau d_{12}(\kappa t,\tau)F^\ast(\tau)\nonumber\\
&+&iG_m|\frac{{\cal E}}{s_m\delta}|^2 \int_0^{\kappa t} d\tau d_{13}(\kappa t,\tau)|F(\tau)|^2e^{is_m\tau}-iG_m|\frac{{\cal E}}{s_m\delta}|^2 \int_0^{\kappa t} d\tau d_{14}(\kappa t,\tau)|F(\tau)|^2 e^{-is_m\tau}\big|^2
\label{c3}
\end{eqnarray}
purely from the coherent drive term, with $F(\tau)=e^{i\delta s_m \tau}-1$.
In Eq. (\ref{c3}) we have also included the contribution from the pure drive action $U_0(t)={\cal T}\exp\{-i\int_0^t d\tau H_e(\tau)\}$. The notation $d_{ij}(t,\tau)=[{\cal T}e^{\int_\tau^t dt' \hat{M}(t')}]_{ij}$ represents a matrix element 
for $i,j=1,2,3,4$. For a sufficiently high drive intensity $E$, the third contribution from the coherent drive term dominates over the two others. 

\begin{figure}[h!]
\vspace{-0cm}
\centering
\epsfig{file=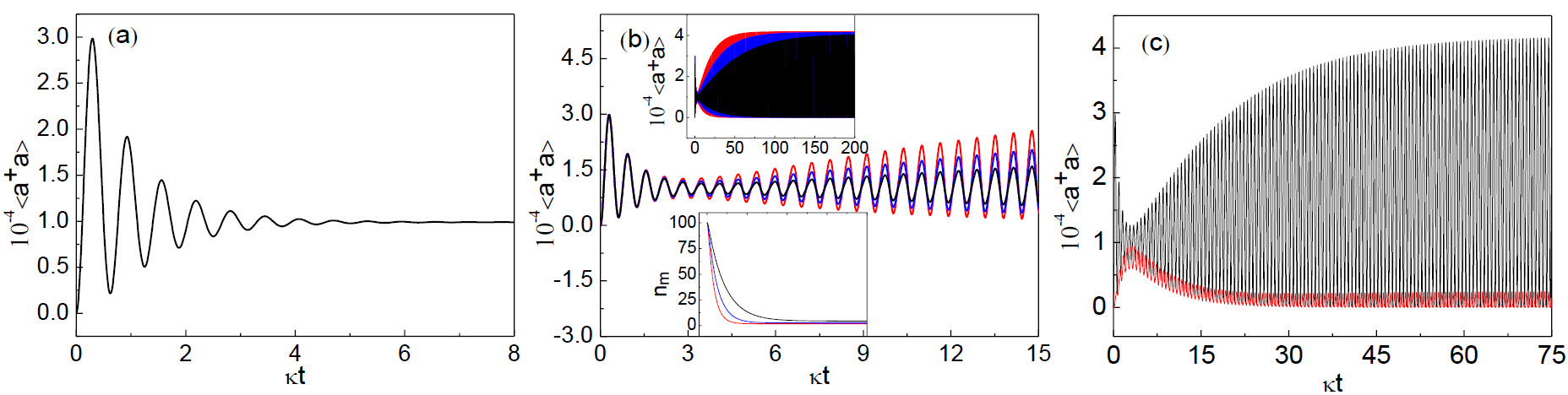,width=1.0\linewidth,clip=} 
{\vspace{-0.2cm}\caption{(a): Photon number evolution without optomechaical coupling ($J=0$). (b): Evolving cavity photon numbers for three different OMSs with $J=0.15$ (black), $J=0.20$ (blue), and $J=0.25$ (red), respectively. The upper and lower inset respectively show the long-time evolutions of the cavity photon numbers and the corresponding thermal phonon numbers, sharing the same horizontal axis range. (c): Mechanism to have the final photon numbers in (b), using the example of $J=0.25$. The red plot is from solving the differential equation Eq. (\ref{dynamics}) only. After including the first term inside the absolute sign in Eq. (\ref{c3}), one has the total photon number as the black plot. 
In these figures we consider a drive with the fixed intensity $E=10^3\kappa$ and the detuning 
$\Delta=\omega_m$, as well as the parameter $\omega_m/\kappa=10$ and the ratio $\gamma_m/\kappa=10^{-3}$. 
The initial thermal occupation of the mechanical mode is $n_{th}=100$. Particularly, the different values of $J$ are realized by choosing the different single-photon optomechanical coupling intensity $g$.}}
\vspace{-0cm}
\end{figure} 

\begin{figure}[h!]
\vspace{-0cm}
\centering
\epsfig{file=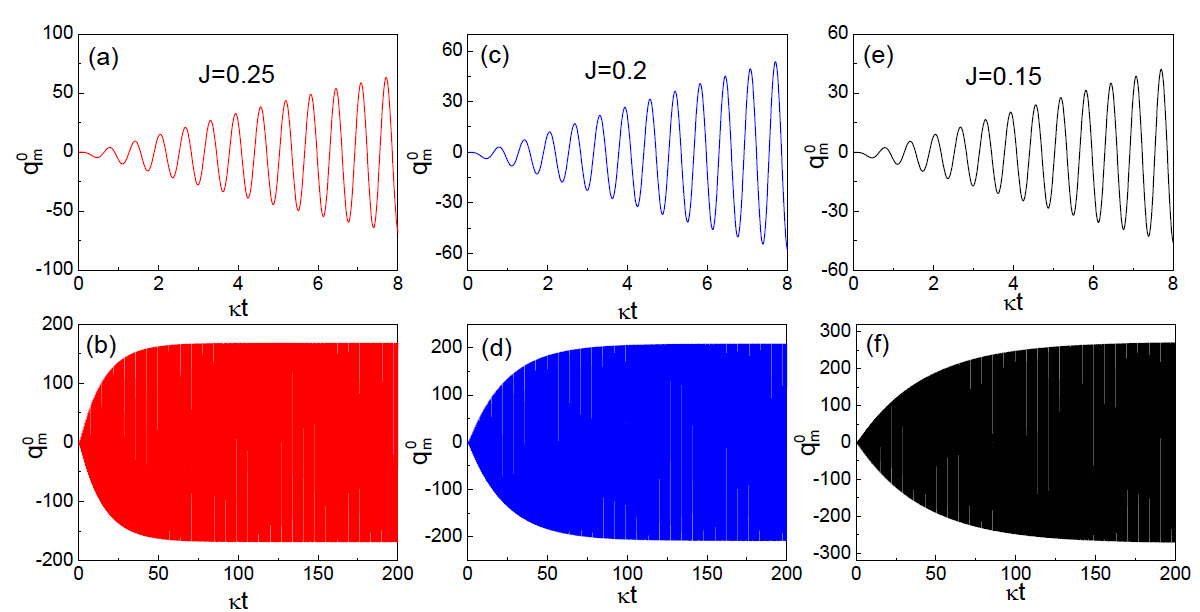,width=0.8\linewidth,clip=} 
{\vspace{-0.2cm}
\caption{Mechanical mode's dimensionless displacements in the situations of Fig. S-1(b). The oscillation frequency is $\omega_m$. The upper and lower column show the short-time and long-time evolutions, respectively. In each situation the coherent drive term $\vec{\lambda}(t)$ alone leads to a steady displacement after a time decided by $J$. After adding the motion of the rotating frame at the frequency $\omega_m$, the finally stabilized displacements will show the stable oscillations. Under the radiation pressure of insignificant intensities as considered here, the mechanical oscillator's amplitude will become larger if it takes a longer time to get to the dynamical stability, to be consistent with the tendency of the stabilized thermal phonon number in Fig. S-1(b)---a lower residual phonon number is achieved by a higher parameter $J$ as the result of dynamical evolution.}}
\vspace{-0.4cm}
\end{figure} 

We here study the cavity photon number evolutions in the weak coupling regime of small effective coupling intensity $J=(g/\omega_m)(E/\kappa)$, as a complement of the discussion on the relatively large $J$ in Figs. 2(a) and 2(c) of the main text. In Fig. S-1(a) about $J=0$, the numerical calculation gives a perfect agreement with the known analytical result of a cavity with fixed size---the photon number tends to a steady value after a period in the order of $1/\kappa$. By intuition, a small $J$ would only lead to perturbative effect, but this is true only at the beginning period of the dynamical evolutions---the plots in Fig. S-1(b) first show a tendency of getting close to the corresponding steady value in Fig. S-1(a), but keep the growing oscillations due to the existence of optomechanical interaction. The finally stabilized photon number evolutions are stably oscillating ones as shown in the upper inset of Fig. S-1(b), in which the photon numbers tend to such a stable dynamical evolution phase together with the corresponding thermal phonon numbers in these cooling processes; compare the corresponding plots in the lower and upper insets of Fig. S-1(b). The photon number curves in the upper inset of Fig. S-1(b) look stuck together because the oscillation period is much shorter than the time scale to reach the finally stable oscillations. 
The corresponding dimensionless displacements $q_m^0(t)= \frac{1}{\sqrt{2}}\langle \hat{b}(t)+\hat{b}^\dagger(t)\rangle$ of the mechanical oscillator for the three different situations in Fig. S-1(b) are plotted in Fig. S-2; they respectively become stable oscillations together with the cavity photon numbers and thermal phonon numbers.
Such stable oscillation looks similar to those of classical OMSs due to the saturated nonlinear effect in the blue-detuned regime (see Sec. VIII in \cite{rev}), but has a different physical origin from that type of classical motion since we are currently concerned with the red-detuned regime. Another interesting phenomenon in the processes is the considerably increased average cavity photon numbers from that of an uncoupled system, when the systems become stabilized. We illustrate its mechanism in Fig. S-1(c), which shows that the increased photon numbers in cavity is due to the constructive interference of two factors, the displacement term irrelevant to the optomechanical coupling [the first term, which is from the unitary action $U_0(t)$, inside the absolute sign in Eq. (\ref{c3})] and the cavity mode amplitude induced by the optomechanical coupling (the summation of the rest terms, which come from the coherent drive term $\vec{\lambda}(t)$ in Eq. (\ref{dynamics}), in the same equation). The latter can be regarded as a relatively small effect for a weakly coupled OMS. Such constructive interference, which is predicted with a completely quantum mechanical treatment for an OMS entering its macroscopic quantum states, indicates that a small coupling $J$ can still affect the cavity photon number considerably.
Enhanced cavity photon numbers in oscillation is a signature of macroscopic quantum states for the OMS.  

There are two intrinsic time scales for reaching the dynamical stability in the cooling of an OMS, $1/\kappa$ for the cavity mode and $1/\gamma_m$ for the mechanical mode. When these two modes are uncoupled ($J=0$), the cavity mode tends to a steady state within a time in the order of $1/\kappa$ 
as in Fig. S-1(a), while the mechanical mode under a thermal equilibrium with the environment satisfies the equation
\begin{eqnarray}
\dot{\hat{b}}&=&-\gamma_m\hat{b}+\sqrt{2\gamma_m}\hat{\xi}_m(t),
\end{eqnarray}
which is from Eq. (\ref{e1}). 
According to the above equation, the mechanical mode takes the time of $1/\gamma_m\gg 1/\kappa$ to damp to zero, a stable value, while the second noise drive term keeps the thermal phonon number $n_{th}$ invariant in this situation. Once these two system modes are coupled ($J\neq 0$), the time scale $t_s$ for the OMS to approach the stable evolution phase will be 
decided by their effective coupling intensity $J$ and be reduced from $1/\gamma_m$ due to the coupling to the cavity mode with a faster damping rate. 
For example, in the limit of $\omega_m/\kappa \rightarrow \infty$, the eigenvalues of the dynamical matrix $\hat{M}$ 
in Eq. (\ref{dynamics}) degenerate to the analytical form $\lambda_{\pm}=\frac{1}{2}(-\kappa-\gamma_m\pm \sqrt{(\kappa-\gamma_m)^2-4J^2\kappa^2})$. The dynamical stability controlled by the two eigenvalues will be realized 
after a long time in the order of $1/\lambda_+\sim 1/\gamma_m$, if the coupling $J$ is very small. This explains the time scale to reach the stable evolution phases in Fig. S-1(b). By the time (in the order of $1/\kappa$) when the uncoupled cavity field in Fig. S-1(a) becomes steady, the coupled system is far from being in dynamical stability, so the cavity mode will keep the unstable oscillation of growing to higher amplitude until the oscillation under the cavity field's 
damping completely stabilizes after much longer time. 
In contrast, the tendency of first dropping to a lower oscillation amplitude (close to the steady state of $J=0$) and then growing to a higher amplitude is less obvious in Fig. 2 of the main text, because at the beginning period the effects of the 
larger values of $J$ are no longer perturbative. Moreover, as the intensity of $J$ becomes still larger, the cavity photon numbers (if they finally stabilize) may exhibit more complicated periodic patterns due to the more obvious manifestation of 
the higher harmonic components $n\omega_m$ ($n>1$) from the dynamical matrix $\hat{M}(t)$ in Eq. (\ref{dynamics}).

A stably oscillating average mechanical mode $\langle \hat{b}(t)\rangle$ in the end, instead of a stopped one, is a major difference from the previous picture. The classical dynamical equation for the cavity mean value, which is due to a finally stabilized mechanical oscillation $\langle \hat{b}(t)\rangle=\beta e^{-i\omega_m t}$ as in Fig. S-2, can be found from 
Eq. (\ref{e2}) as
\begin{eqnarray}
\dot{\alpha}&=&-\kappa\alpha-\big(i\Delta-ig(\beta e^{-i\omega_m t}+\beta e^{i\omega_m t})\big)\alpha+E.
\label{cl-solution}
\end{eqnarray}
The solution to this equation is the series 
$\alpha(t)=e^{i\phi(t)}\sum_{n=-\infty}^\infty\alpha_n e^{in\omega_m t}$ with 
\begin{eqnarray}
\alpha_n=E\frac{J_n(-\frac{g\beta}{\omega_m})}{i(n\omega_m+\Delta)+\kappa},
\label{series}
\end{eqnarray}
where $J_n(x)$ is the Bessel function of the first kind and $\phi(t)$ is a global phase \cite{supp2, supp0}. 
The result $|\alpha_0 +\alpha_{-1}e^{-i\omega_m t}+\alpha_{1}e^{i\omega_m t}|^2$, which gives higher cavity photon numbers than those of uncoupled systems ($J=0$) due to the interference with the extra sidebands, can qualitatively explain the stabilized photon numbers. For example, because the finally stabilized factors $g\beta$ are close to 
a same value for the three different oscillations in Fig. S-2, the corresponding photon numbers that are to be stabilized will become close to one another as shown by the upper inset in Fig. S-1(b).

Static expectation values of the cavity and mechanical modes, i.e. $\frac{d}{dt}\langle\hat{a}(t)\rangle=\frac{d}{dt}\langle\hat{b}(t)\rangle=0$, for the final state of a cooling process were assumed in the previous studies; see, e.g. the vanishing amplitude $\beta \rightarrow 0$ 
of the mechanical oscillation as described in Sec. VIII of Ref. \cite{rev}. According to the classical mean field $\alpha(t)$ determined with the above Eqs. (\ref{cl-solution}) and (\ref{series}), such vanishing mechanical oscillation amplitude will lead to a cavity photon number indifferent to that of an uncoupled cavity ($J=0$). 
If the mechanical oscillator is assumed to stop moving at a nonzero displacement $\langle \hat{q}^0_m\rangle=\bar {x}_m$, 
the cavity field amplitude [only from the contribution of the term containing $J_0(-\frac{g\beta}{\omega_m})$] will become $\alpha_0=E/\{i(\Delta-g\bar{x}_m)+\kappa\}$, an exact steady-state solution to be reached within a time $t\sim 1/\kappa$ [similar to the evolution in Fig. S-1(a)]. The static picture thus implies the separate dynamical evolutions of the cavity and mechanical modes, which are very different from the illustrations in Fig. S-1(b).  

The cooling of a mechanical oscillator starts from its thermal state with the Wigner function 
$W(q_m, p_m)=\frac{1}{\pi(1+2n_{th})}\exp\{-\frac{1}{(1+2n_{th})}(q_m^2+p_m^2)\}$. 
A perfect cooling is to the ground state $|0\rangle_m$ of the mechanical oscillator with its Wigner function 
$W(q_m, p_m)=\frac{1}{\pi}\exp\{-(q_m^2+p_m^2)\}$, having the associated thermal phonon number reduced from $n_{th}$ to $0$. Due to the inevitable motion of the mechanical mode under the radiation pressure for cooling, in such perfect cooling its Wigner function will actually become $W(q_m, p_m)=\frac{1}{\pi}\exp\{-(q_m-q_m^0(t))^2+(p_m-p_m^0(t))^2)\}$, which is that of a coherent state with the displacements $q_m^0(t)$, $p_m^0(t)$ in the phase space. An arbitrary time-dependent function $q_m^0(t)$ or $p_m^0(t)$ does not affect the cooling result, since the purity of this target coherent state is irrelevant to the average position and average momentum of the oscillator. Actually, cooling a mechanical oscillator is to realize its pure macroscopic quantum states, no need to let the cavity and mechanical mode be frozen in the phase space. 
More discussion on the picture can be found in the next section.

\subsection{IV. FURTHER DISCUSSION ON THE EFFECTS OF THE NOISE DRIVES}

\renewcommand{\theequation}{S-IV-\arabic{equation}}
\setcounter{equation}{0}
\begin{figure}[h!]
\vspace{-0cm}
\centering
\epsfig{file=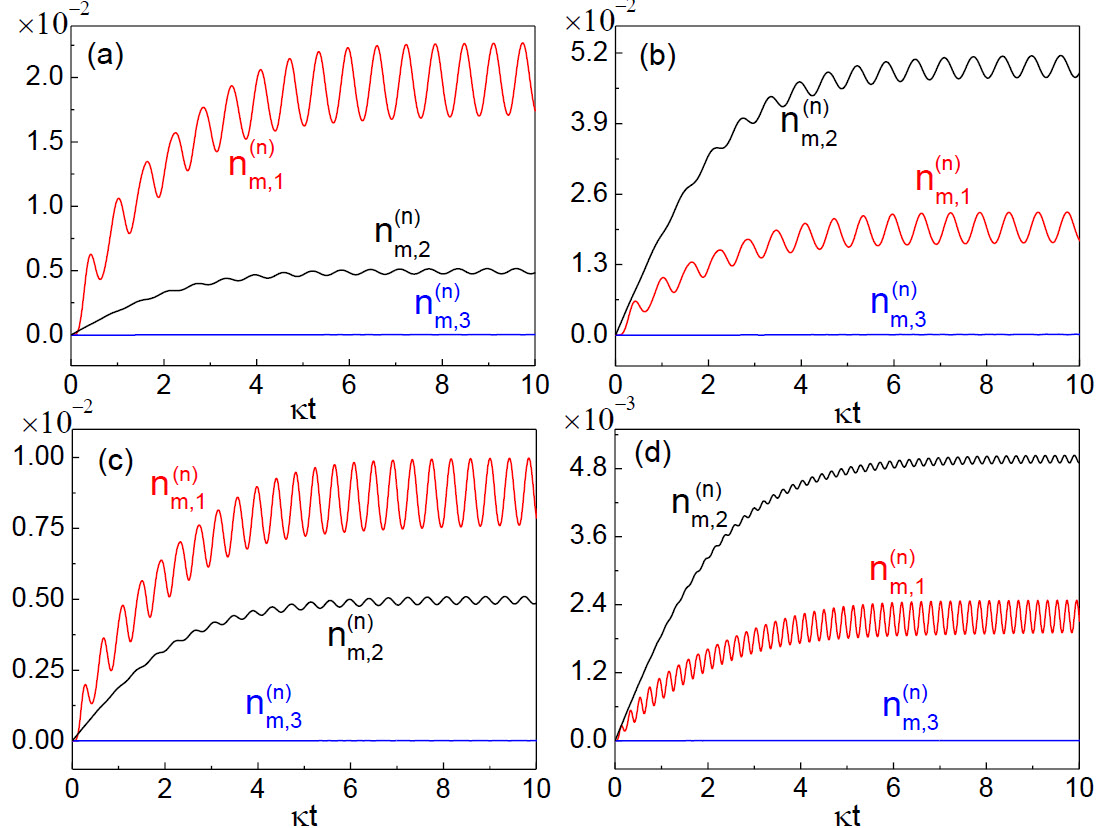,width=0.55\linewidth,clip=} 
{\vspace{-0.2cm}\caption{Contributions to thermal phonon number from different noise drives. Here we set $J=0.5$ for the OMSs with $g=10^{-5}\kappa$ and $\gamma_m=10^{-3}\kappa$. The drive detuning is chosen at $\Delta=\omega_m$.
(a) $\omega_m/\kappa=10$ and $n_{th}=1$; (b) $\omega_m/\kappa=10$ and $n_{th}=10$; (c) $\omega_m/\kappa=15$ and $n_{th}=1$; (d) $\omega_m/\kappa=30$ and $n_{th}=1$.}}
\vspace{-0.4cm}
\end{figure}

In the calculation of the thermal phonon numbers, we have excluded the coherent drive term 
$\vec{\lambda}(t)$'s contribution.
The displacement Hamiltonian $H_{displace}(t)$ in Eq. (\ref{Hi}) certainly contributes to the total occupation 
$\langle \hat{b}^\dagger\hat{b}\rangle$ of the mechanical mode, which will not drop to a lower level throughout a cooling process as indicated by the overall mechanical motion like those in Fig. S-2. Even if the mechanical oscillator were stopped, the resulting nonzero displacement due to the radiation pressure would still significantly contribute to $\langle \hat{b}^\dagger\hat{b}\rangle$ different from our concerned thermal phonon number. 
In the previous linearization of an OMS's dynamical equations by shifting cavity mode by its steady-state value $\alpha$, such effect exhibits as a displacement term $ig|\alpha|^2$ from Eq. (\ref{e2}). After excluding the coherent drive term's contribution, the result of a cooling process is determined by the coexisting action of the noise drives, as the finally stabilized thermal phonon number in Eq. (7) of the main text. 

The noise contributions from different sources can be identified as the three different terms in Eq. (7) of the main text. They are, respectively, $n_{m,1}^{(n)}$ from the cavity noise drive via squeezing (SQ) type coupling, $n_{m,2}^{(n)}$ from the mechanical reservoir noise drive via BS type coupling, and $n_{m,3}^{(n)}$ from the mechanical noise drive via SQ type coupling. The first contribution, which exists even at the zero temperature due to the remnant squeezing effect, 
can be lowered with increased sideband resolution $\omega_m/\kappa$ as seen from Figs. S-3(c) and S-3(d). The cooling limit when the parameter $\omega_m/\kappa$ tends to infinity comes from the second contribution proportional to equilibrium thermal occupation $n_{th}$. Due to the small ratio $\gamma_m/\kappa$ and a BS resonant coupling used for cooling, the third contribution can be totally neglected.

Such identification of different quantum noise origins for stabilized thermal phonon number is not so explicit with the other approaches. The earliest approach to quantum optomechanical cooling borrows the picture for cooling of trapped ions 
and for cavity assisted laser cooling of atomic and molecular motion. Similar to that picture, the mechanical oscillator 
is modeled as a quantum noise spectrometer responding to the cavity field \cite{supp3}, 
which is phenomenologically treated as a reservoir, to derive an effective master equation 
 \begin{eqnarray}
\dot{\rho}_m(t)&=&-i[\omega_m\hat{b}^\dagger\hat{b},\rho_m(t)]-(\gamma_m (n_{th}+1)+A_{-})\big(\hat{b}^\dagger\hat{b}\rho_m(t)+\rho_m(t)\hat{b}^\dagger\hat{b}-2\hat{b}\rho_m(t)
\hat{b}^\dagger\big)\nonumber\\
&-&(\gamma_m n_{th}+A_+)\big(\hat{b}\hat{b}^\dagger\rho_m(t)+\rho_m(t)\hat{b}\hat{b}^\dagger-2\hat{b}^\dagger\rho_m(t)\hat{b}\big)
\label{reduced}
\end{eqnarray}
for the mechanical oscillator's quantum state $\rho_m$.
The transition rate $A_{\pm}=g^2|\alpha|^2\frac{2\kappa}{\kappa^2+(\Delta\pm \omega_m)^2}$ in the equation is decided by a time-independent amplitude $\alpha$ of the cavity mean field \cite{c1,c2}, 
which is approximated with a zeroth-order term in the solution given by Eq. (\ref{series}) or 
a steady-state field in a cavity of fixed size as in Fig. S-1(a). The corresponding equation of motion for the averaged mechanical mode $\langle \hat{b}(t)\rangle=\mbox{Tr}\{\hat{b}\rho_m(t)\}$ reads
\begin{eqnarray}
\frac{d}{dt}\langle \hat{b}\rangle=-i\omega_m \langle \hat{b}\rangle-\{\gamma_m+(A_{-}-A_+)\}\langle \hat{b}\rangle,
\label{b-mode}
\end{eqnarray} 
implying that the mechanical oscillator will gradually become motionless after a time $t\sim 1/(\gamma_m+\Gamma_{opt})$ where $\Gamma_{opt}=A_{-}-A_+$. In the weak coupling regime with $\gamma_m+\Gamma_{opt} \ll \kappa$, 
this will lag far behind the time $t\sim 1/\kappa$ for the cavity field to reach the assumed steady-state value $\alpha=\langle \hat{a}\rangle=E/(i\Delta+\kappa)$. The Heisenberg-Langevin equations (such as those for the coupled cavity field fluctuation and the mechanical mode in \cite{rev}) derived previously with the steady-state value $\alpha$ also reflect this notion of asynchronous evolution, which is an approximation for the actual quantum optomechanical cooling processes. 
For instance, shortly after the moment of turning on the driving laser, the mechanical oscillator 
in a thermal state [the initial state $\rho_m(0)$ for Eq. (\ref{reduced})] cannot immediately 
have the transition rate $A_{\pm}$ proportional to the steady-state value $|\alpha|^2$, because the cavity field has not evolved to that assumed steady state yet. Even the evolution in Fig. S-1(a) will take time to 
become steady, though the time scale is much shorter than that for the process depicted 
by Eq. (\ref{b-mode}). The quantum state $\rho_m$ of the mechanical mode will be definitely changed under the coupling with a developing cavity field before it arrives at any possible stable evolution or steady state, so that the details of the transient period become relevant to the cooling result. 

In terms of the motion of two coupled harmonic oscillators (as described by the averaged Eqs. (\ref{e1}) and 
(\ref{e2}) for a weakly coupled system $g/\omega_m\ll 1$) and the dynamics of the associated system-mode fluctuations, 
the pictures for the different treatments of quantum optomechanical cooling are summarized as below. Here the described motions are from the same static initial condition for the oscillators, i.e. the expectation values $\langle \hat{a}\rangle=\langle \hat{b}\rangle=0$ determined from 
the cavity vacuum state and mechanical thermal state at $t=0$. 

\noindent ---(A) the effective approach based on steady states:

$\bullet$ Oscillator A representing the cavity field first stops at $t\sim 1/\kappa$, and then oscillator B representing 
the mechanical vibration will stop after $t\sim 1/(\gamma_m+\Gamma_{opt})$ to end a cooling process.

$\bullet$ An alternative picture is to consider both of the fluctuations 
$\delta\hat{a}=\hat{a}-\alpha$ and $\delta\hat{b}=\hat{b}-\beta$, where the time-independent steady cavity field amplitude $\alpha$ and steady mechanical displacement $\beta$ are found from the averaged form of Eq. (\ref{e2}). 
These fluctuations have been used to interpret optomechanical cooling \cite{ck} and optomechanical entanglement \cite{supp4}. According to the linearized dynamical equations to govern the evolutions of the fluctuations driven by the noises, which are derived with the steady-state values $\alpha$ and $\beta$, the relevant physical quantities (such as $\langle \delta \hat{x}_m^2\rangle$ and $\langle \delta \hat{p}_m^2\rangle$) will inevitably evolve to time-independent values in the end.

\noindent ---(B) our dynamical approach:

$\bullet$ The coupled oscillators A and B start moving due to an external force. Then they will gradually turn into stable oscillations together, unlike a process to trap the mechanical oscillator's motion to the ground state in an external 
simple harmonic potential so that $\langle\hat{x}_m \rangle=\langle \hat{p}_m\rangle=0$. 
All finally realized physical quantities in a cooling process are determined in the transient period toward the 
dynamical stability of oscillators A and B.

$\bullet$ In terms of the fluctuation $\delta\hat{b}(t)=\hat{b}(t)-\langle \hat{b}(t)\rangle$ around 
the time-dependent expectation value $\langle \hat{b}(t)\rangle$ that can give the motions in Fig. S-2, our concerned thermal occupation  $\langle\hat{b}^\dagger\hat{b}\rangle-|\langle\hat{b}\rangle|^2=\langle (\hat{b}^\dagger-\langle \hat{b}\rangle^\ast)(\hat{b}-\langle \hat{b}\rangle)\rangle$ can also be expressed as $\langle \delta \hat{b}^\dagger\delta \hat{b}\rangle$. The contributions from the different noise drive terms to the fluctuation $\delta \hat{b}(t)$ will lead to the stable oscillations as those in Fig. S-3 (when viewed with small scales), so there will not be exactly time-independent steady values. In a cooling process, the quantity $\langle \delta \hat{b}^\dagger\delta \hat{b}(t)\rangle$ will be lowered, 
while the purity of the macroscopic quantum state $\rho_m(t)$ of the mechanical mode will become higher. Note that the thermal phonon number $\langle \delta \hat{b}^\dagger\delta \hat{b}(t)\rangle$ also includes the contribution 
from the system-operator part as Eq. (6) in the main text, which will asymptotically tend to zero by the time 
$t\sim t_s$ to reach the dynamical stability. 

Two qualitative differences from our approach will still exist even when the system dynamics is linearized by shifting $\hat{a}(t)\rightarrow \hat{a}(t)+\alpha(t)$ with a time-dependent mean field $\alpha(t)$. (1) The mean-field solution $\alpha(t)$ of the nonlinear dynamical equations can become undetermined, for example, in the regimes around bistability. 
(2) The magnitude of $\alpha(t)$ should be much higher than that of the cavity field fluctuation so that a linearized Hamiltonian like that in Eq. (\ref{lnh}) [with the steady-state value $\alpha$ in the equation replaced by an evolving one $\alpha(t)$] can be obtained. The difference in the linearized Hamiltonians indicates that, even if such simulation with 
an evolving $\alpha(t)$ could be performed, the simulated evolution of thermal occupation or thermal phonon number will not be similar to the prediction by our quantum dynamical approach.

\begin{figure}[t!]
\vspace{-0.3cm}
\centering
\epsfig{file=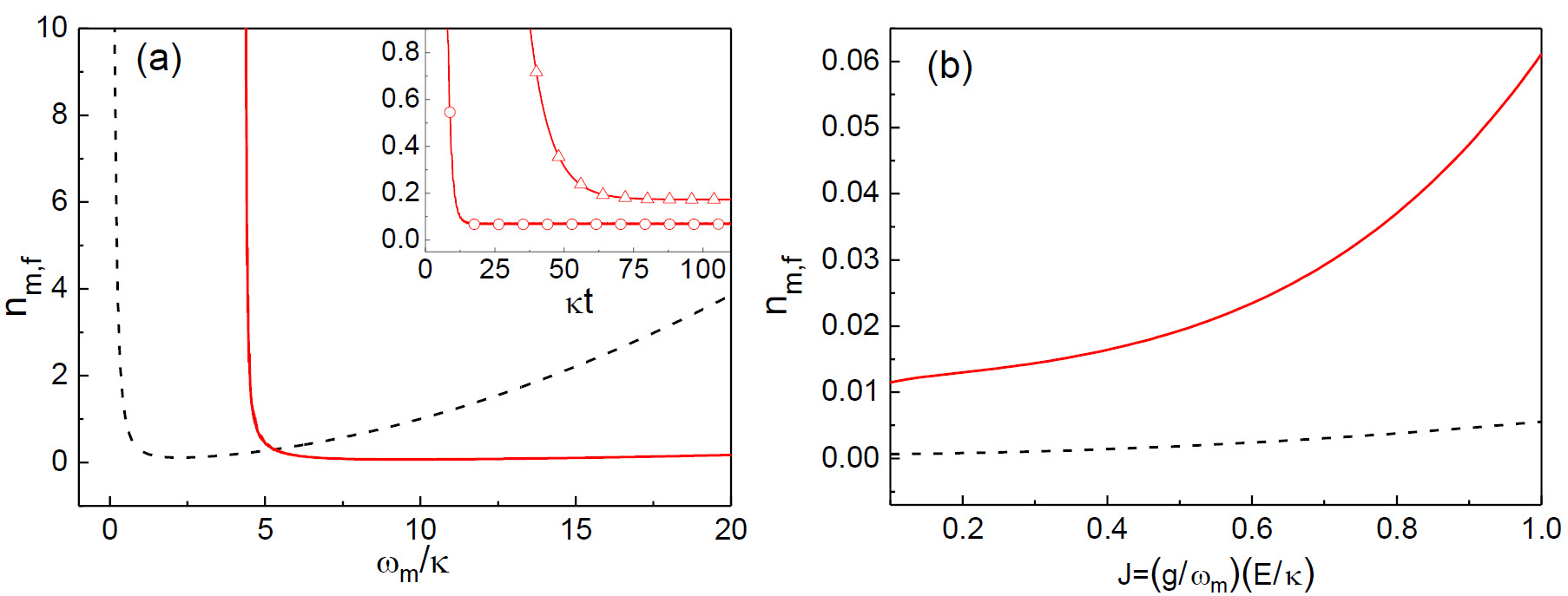,width=0.7\linewidth,clip=} 
{\vspace{-0.2cm}\caption{(a) Comparison of the residual thermal phonon numbers achieved in the weak coupling regime, according to the different approaches. The solid curve is our prediction, and the dashed one is found with Eq. (\ref{weak-1}). Here we use one set of the system parameters in \cite{c2}: $n_{th}=100$, $\Delta=\omega_m$, $\gamma_m/\kappa=10^{-4}$, $(E/\kappa)^2=10^5$, and $\Gamma_{opt}(|\alpha|^2=1,\omega_m\rightarrow \infty)/\gamma_m=0.1$ (equivalent to $(g/\kappa)^2=\frac{1}{2}10^{-5}$). The inserted figure shows our prediction of the evolutions of thermal phonon number for $\omega_m/\kappa=10$ (circle) 
and $\omega_m/\kappa=20$ (triangle). (b) Comparison of our predicted pure cavity noise contribution at $T=0$ ($n_{th}=0$) 
and $\Delta=\omega_m$ with a previous prediction made for the strong coupling regime. 
The dashed curve is obtained according to Eq. (\ref{p}). For this comparison, we use the different parameters---$g=10^{-5}\kappa$, $\gamma_m/\kappa=10^{-3}$, and $\omega_m/\kappa=10$.}}
\vspace{-0cm}
\end{figure}

The results of cooling according to the distinct pictures can have big difference. 
From the steady-state solution $\dot{\rho}_m(t)=0$ of Eq. (\ref{reduced}), 
the thermal occupation at a final steady state will be found as 
\begin{eqnarray}
n_{m,f}=\frac{\Gamma_{opt}n_m^0+2\gamma_m n_{th}}{\Gamma_{opt}+2\gamma_m},
\label{weak-1}
\end{eqnarray} 
where
\begin{eqnarray}
\Gamma_{opt}=\frac{2g^2|\alpha|^2}{\kappa}\frac{1}{1+(\frac{\kappa}{2\omega_m})^2}
\label{rate}
\end{eqnarray} 
and $n^0_m=(\kappa/2\omega_m)^2$, $|\alpha|^2=(E/\kappa)^2/(1+(\Delta/\kappa)^2)$ 
\cite{c2}. The coefficients in these formulas have slight variations from those 
in the previous literature, due to our notation for the damping terms ($-\kappa \hat{a}$, $-\gamma_m \hat{b}$ in the dynamical equations) as compared with the corresponding ones $-\frac{\kappa}{2}\hat{a}$ and $-\frac{\gamma_m}{2} \hat{b}$ 
in \cite{c1,c2}, and our definition of detuning $\Delta$ also differs by an opposite sign. 
Similar steady-state occupation can also be found for other systems (see, e.g. \cite{supp5}).
In Eq. (\ref{weak-1}) the first part from the cavity noise's back-action is almost independent of a cooling rate $\Gamma_{opt}\gg \gamma_m$ and is thus beyond the control by the drive power. 
A contrasting feature of our results is that the thermal phonon number 
can be totally controlled by a parameter $J$ proportional to the drive intensity $E$, 
since each contribution to the finally stabilized thermal occupation shown in Fig. S-3 
is decided by how soon its evolution becomes stable. 

The explicit comparison between a result from Eq. (\ref{weak-1}) and the prediction by our dynamical approach is shown in Fig. S-4(a). Given a sufficiently small ratio $\gamma_m/\kappa$ or a sufficiently high quality factor of the mechanical oscillator as in the figure, our predicted thermal phonon number to be stabilized changes slightly with increased 
sideband resolution $\omega_m/\kappa$, in contrast to an obvious optimum sideband resolution $\omega_m/\kappa$ found with 
Eq. (\ref{weak-1}). Another obvious difference is in an assumed situation of initial zero temperature, where the residual phonon number is known as pure quantum limit or quantum back-action limit. 
This contribution as the first term in Eq. (7) of the main text increases significantly with the parameter $J$. 
A previous development from the above-mentioned treatment, on the other hand, predicts the value in the strong coupling regime as \cite{c3} 
\begin{eqnarray}
n_{m,f}=\frac{\kappa^2}{4\omega_m^2}+\frac{g^2|\alpha|^2}{2\omega_m^2},
\label{p}
\end{eqnarray}
where a zero thermal occupation for the cavity reservoir is assumed, 
and the different coefficients from those in \cite{c3} are due to the same reason for Eqs. (\ref{weak-1}) and (\ref{rate}). Our result, the solid curve in Fig. S-4(b), indicates a much more significant effect of the noise from the cavity reservoir.
 
\begin{figure}[b!]
\vspace{-0.3cm}
\centering
\epsfig{file=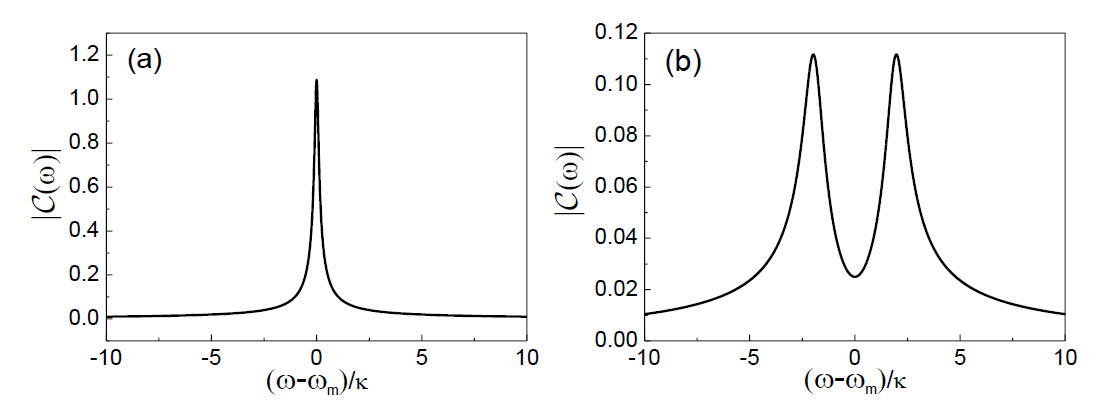,width=0.75\linewidth,clip=} 
{\vspace{-0.2cm}\caption{Noise spectrum of the cavity field in the weak and strong coupling regime, respectively. (a) $J=0.3$; (b) $J=2.0$.
The other system parameters are $\gamma_m/\kappa=10^{-3}$, $n_{th}=100$, and $\omega_m/\kappa\rightarrow \infty$. Here the horizontal axis has been shifted with respect to $\Delta=\omega_m$.}}
\vspace{-0cm}
\end{figure}

The last issue is how to know the cooling result. The correlation function 
$\langle \hat{i}(t+\tau)\hat{i}(t)\rangle-\langle \hat{i}(t+\tau)\rangle\langle \hat{i}(t)\rangle$ \cite{supp6}, where $\hat{i}(t)=\hat{a}_{out}(t)+\hat{a}^\dagger_{out}(t)\rangle$ and $\hat{a}_{out}(t)=E(t)/\sqrt{\kappa}+\sqrt{\kappa}\hat{a}(t)$, after the time $t \sim t_s$ for reaching the dynamically stable phase is in one-to-one correspondence with the finally achieved thermal occupation or thermal phonon number.
It primarily consists of the following correlation of the cavity mode:
\begin{eqnarray}
{\cal C}(\tau)& = & \langle \hat{a}^\dagger(t+\tau)\hat{a}(t)\rangle- \langle \hat{a}^\dagger(t+\tau)\rangle \langle \hat{a}(t)\rangle.
\end{eqnarray}
The Fourier transform of this function, ${\cal C}(\omega)=\int d\tau {\cal C}(\tau)e^{i\omega \tau}$, constitutes the main component in a measured noise spectrum. Such correlation function is only determined by the noise drive term in Eq. (\ref{dynamics}). If the contribution from the coherent drive term in the equation is also included, the Fourier transform of $\langle \hat{a}^\dagger(t+\tau)\hat{a}(t)\rangle$ will have an extra delta-function term from the finally stable cavity field oscillation as the background (see, e.g. \cite{c1}). The noise spectra for the cooling processes taking place close to the limiting boundary (the dashed curve in Fig. 4 of the main text) are illustrated in Fig. S-5. The single-peak one in Fig. S-5(a) reflects the thermometry results in the past experiments.
\end{widetext}

\end{document}